\documentclass[letterpaper,twocolumn,prl,
aps,showpacs,superscriptaddress,floatfix]{revtex4-1}
\usepackage{microtype}
\usepackage[latin1]{inputenc}
\usepackage{bm}
\usepackage[usenames]{color}
\usepackage{multirow}
\usepackage{amssymb}
\usepackage{amsbsy}
\usepackage{mathtools}
\usepackage{amsmath}
\usepackage{stmaryrd}
\usepackage{graphicx}
\usepackage{epsfig}
\usepackage{placeins}
\usepackage{ulem}
\usepackage{nicefrac}
\usepackage{bbold}
\usepackage{braket}
\usepackage{blindtext}
\usepackage[colorlinks,linkcolor=blue,citecolor=blue,urlcolor=blue]{hyperref}
\usepackage[capitalize]{cleveref}

\makeatletter
 
\begin{document}
\title{Transport and entanglement growth in long-range 
random Clifford circuits}

\author{Jonas Richter}
\affiliation{Department of Physics and Astronomy, University College London, 
Gower Street, London, WC1E 6BT, UK}
\affiliation{Department of Physics, Stanford University, Stanford, CA 94305, 
USA}
\affiliation{Institut f\"ur Theoretische Physik, Leibniz Universit\"at 
Hannover, Appelstra\ss e 2, 30167 Hannover, Germany}

\author{Oliver Lunt}
\affiliation{Department of Physics, King's College London, Strand, London, WC2R 2LS, UK}
\affiliation{School of Physics and Astronomy, University of Birmingham, 
Birmingham, B15 2TT, UK}

\author{Arijeet Pal}
\affiliation{Department of Physics and Astronomy, University College London, 
Gower Street, London, WC1E 6BT, UK}

\date{\today}

%---------------------------------------------------------------------------------------------------
\begin{abstract}
Conservation laws can constrain entanglement 
dynamics in isolated quantum systems, manifest in a 
slowdown of higher R\'enyi entropies. Here, we explore 
this phenomenon in a class of long-range random Clifford circuits with U$(1)$ 
symmetry where transport can be tuned from diffusive to superdiffusive.
We unveil that  
the different hydrodynamic regimes reflect themselves in the  
asymptotic entanglement growth according to $S(t) \propto t^{1/z}$, where the  
dynamical transport exponent $z$ depends on the probability $\propto 
r^{-\alpha}$ of gates spanning a distance $r$.
For sufficiently small $\alpha$, we show that the 
presence of hydrodynamic modes becomes irrelevant such that $S(t)$ behaves 
similarly in circuits with and without conservation law. We 
explain our findings in terms of the inhibited 
operator spreading in U$(1)$-symmetric Clifford circuits, 
where the emerging light cones can be understood in the context of classical 
L\'evy flights. Our work sheds light on the connections 
between Clifford circuits and more generic 
many-body quantum dynamics.
\end{abstract}

\maketitle
%-------------------------------------------------------------------------------
%--------------------

{\it Introduction.--}
Fundamental questions on the origin of quantum 
statistical mechanics have experienced a renaissance in recent 
years \cite{dalessio2016, nandkishore2015, Bertini2021}, with 
experiments being able to probe chaos and information scrambling \cite{Li2017, 
Gaerttner2017, Landsman2019, 
Blok2021}.
While much progress has been made due to 
sophisticated numerical  
methods (e.g., \cite{Haegeman2011, Paeckel2019, 
Rakovszky2022, White2018, Wurtz2018,Heitmann2020}), ideas 
from quantum information provide a useful lens on
quantum dynamics far from equilibrium. In particular, suitable random-circuit 
models capture aspects of generic quantum systems 
\cite{Nahum2017, Keyserlingk2018, Nahum2018, Brandao2021}, 
including settings with conservation 
laws and constraints \cite{Khemani2018, Rakovszky2018, Moudgalya2021}, as well 
as dual-unitary 
\cite{Bertini2019, Claeys2021}, time-periodic \cite{Bertini2018, Chan2018}, or 
nonunitary dynamics \cite{Skinner2019, Potter2021}.
Random circuits are particularly attractive 
in view of today's noisy intermediate-scale quantum devices 
\cite{Preskill2018, Richter2021, Lunt2021}, with  
applications in achieving
a quantum computational advantage \cite{Arute2019} and 
exploring operator entanglement \cite{Mi2021}.
\begin{figure}[tb]
 \centering
 \includegraphics[width=0.95\columnwidth]{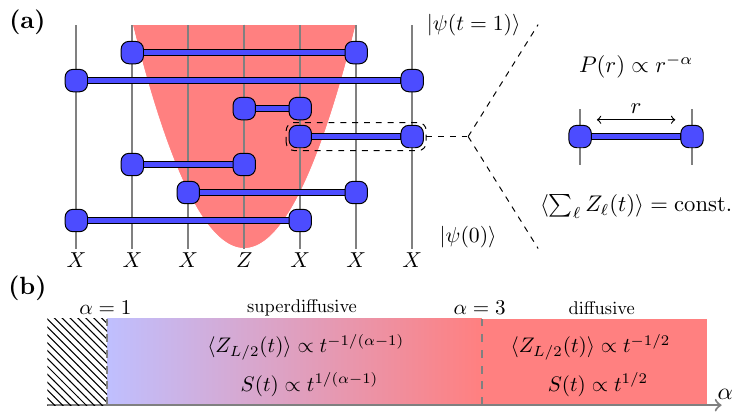}
 \caption{{\bf 
(a)} Two-qubit Clifford  
gates of range $r$ occur with probability $P(r) \propto r^{-\alpha}$ and  
conserve the total Pauli-$Z$ 
component \cite{SuppMat}. {\bf (b)} By tuning 
$\alpha>1$, different hydrodynamic regimes with dynamical exponent $z$ [Eq.\ 
\eqref{Eq::Levy}] emerge, manifest in the tails of 
the circuit-averaged expectation value $\langle 
Z_{L/2}(t)\rangle 
\propto t^{-1/z}$. Entanglement 
saturates approximately on a time scale $\propto L^z$, 
implying that it asymptotically mirrors the transport 
behavior, $S(t) \propto 
t^{1/z}$.
 }
 \label{Fig::Scheme}
\end{figure}

In case of chaotic quantum systems with short-ranged interactions,  
conservation laws give rise to hydrodynamic 
modes that typically decay diffusively \cite{Lux2014, 
Bohrdt2017, Richter2018, Richter2019}, while entanglement is 
expected to 
grow ballistically \cite{Kim2013}. Remarkably, recent work unveiled that 
this picture is incomplete and that transport and entanglement 
are
intimately connected \cite{Rakovszky2019, Huang2020,Znidaric2020, Rakovszky2021, 
Zhou2020_2}. 
Specifically, diffusive transport can 
constrain higher 
R\'enyi entropies to increase 
diffusively \cite{Rakovszky2019}, 
\begin{equation}
 S_{n>1}(t) \propto \sqrt{t}\ ,\quad \text{where}\ S_n = \log_2 
\text{tr}[\rho_A^n]/(1-n), 
\end{equation}
with $\rho_A = \text{tr}_B \ket{\psi(t)}\bra{\psi(t)}$ denoting the reduced  
density matrix for a bipartition into subsystems $A$ and $B$, and 
$\ket{\psi(t)}$ is the state of the system. 
In 
contrast, the von 
Neumann entropy $S_1 = -\text{tr}[\rho_A \log_2 \rho_A]$ grows 
linearly as usual, $S_1(t)\propto t$. In this Letter, we 
demonstrate that constrained entanglement dynamics occurs 
more generically also for other transport types, and can 
be 
readily explored in U$(1)$-symmetric long-range Clifford 
circuits [Fig.\ \ref{Fig::Scheme}~(a)]. Depending on the probability $\propto 
r^{-\alpha}$ of 
gates spanning a distance $r$, the emerging transport can be 
tuned from diffusive to superdiffusive.
These circuits can be seen as 
minimal models to describe the scrambling dynamics of long-range Hamiltonian 
systems. Specifically, it was found in \cite{Zhou2020, Block2022} that the 
light-cone spreading in such circuits is very similar to the dynamics 
generated by Hamiltonians with interactions decaying as $\propto 
r^{-{\alpha^\prime}}$, where $\alpha' = 
\alpha/2$.  While 
Clifford gates are insufficient for universal quantum 
computation, they 
form unitary 2-designs \cite{Harrow2009} (3-designs for qubits 
\cite{Webb2015}), such that circuit averages of 
certain quantities, e.g., out-of-time-ordered correlators, 
coincide with Haar averages over the full unitary group \cite{Nahum2018, 
Keyserlingk2018}. Clifford circuits can thus be useful to study
aspects of 
more generic quantum dynamics. 

Long-range interactions are ubiquitous in nature,  
including dipolar or van der Waals interactions \cite{Saffman2010}, 
experimentally realized in various platforms 
\cite{Porras2004, Jurceviv2014, Smith2016, Richerme2014, Bernien2017, 
Periwal2021, Joshi2021}. In 
contrast to short-range models, where 
Lieb-Robinson bounds confine correlations to a linear ``light 
cone'' \cite{Lieb1972}, 
long-range interactions may lead to faster information 
propagation \cite{Lashkari2013, Avellino2006}. Much effort has been 
invested to tighten Lieb-Robinson-like bounds for power-law 
interacting models \cite{Hastings2006, Eisert2013, Hauke2013, Foss-Feig2015, 
Tran2019, Luitz2019, Colmenarez2020, Zhou2019, 
Zhou2020, Kuwahara2020, Tran2021, Chen2021}, and to 
study transport and entanglement dynamics
\cite{Kloss2019, 
Schuckert2020, Cevolani2016, Schachenmayer2013, Pappalardi2018, Lerose2020,  
Kuwahara2021, 
Block2022, 
Minato2022, Mueller2022,Hashizume2022}. For chaotic systems in $d$ dimensions, 
it was argued that linear light cones arise for $\alpha' > d 
+ 1/2$ with properties similar to short-range models, 
while power-law or logarithmic bounds emerge for $d/2 < \alpha' < d+1/2$ 
\cite{Zhou2019, Zhou2020}. 
For $\alpha' < d/2$, locality breaks down and information 
propagation becomes essentially instantaneous~\cite{Bachelard2013}. 

From a numerical point of view, long-range systems are 
challenging due to quick entanglement 
generation and strong finite-size effects \cite{Zaletel2015}. In contrast, the 
random
Clifford 
circuits considered here can be simulated efficiently even for large 
systems.  
Summarizing our main results, we unveil a direct 
correspondence between transport and entanglement, 
with 
entanglement saturating on a time scale $t_\text{sat} 
\propto L^z$ implying an asymptotic scaling $S(t) \propto 
t^{1/z}$, where $z$ is the dynamical transport exponent [Fig.\ 
\ref{Fig::Scheme}~(b)]. We explain this finding in 
terms of 
the inhibited operator spreading in U$(1)$-symmetric Clifford 
circuits, leading to narrower light cones compared to 
circuits without conservation law. Moreover, we demonstrate that the 
constraint on $S(t)$ becomes insignificant once the dynamical exponent for 
transport reaches $z \approx 1$.

{\it Clifford circuits with symmetry.--}
Clifford circuits are of major interest in quantum information  
\cite{Nielsen2000}, including error correction and 
randomized benchmarking \cite{Knill2008, Magesan2011}. In the context of 
quantum dynamics, they recently gained popularity to study 
measurement-induced entanglement transitions (e.g., \cite{Block2022, 
Li2018, Gullans2020, Sharma2022, Lavasani2021, Lunt2021_2}) as their efficient 
simulability allows to access large system sizes \cite{Aaronson2004, 
Anders2006}. The key idea is to exploit the stabilizer 
formalism \cite{Nielsen2000, Gottesman1998}, where a state $\ket{\psi}$ 
on $L$ qubits can be uniquely defined by $L$ operators ${\cal O}_i$, i.e.,  
${\cal O}_i \ket{\psi} = \ket{\psi}$, where ${\cal O}_i = 
X_1^{\nu_1^i} Z_1^{\mu_1^i} \cdots X_\ell^{\nu_\ell^i}Z_\ell^{\mu_\ell^i} \cdots 
X_L^{\nu_L^i} Z_L^{\mu_L^i}$ are $L$-site Pauli strings and $\nu_\ell^i, 
\mu_\ell^i = \lbrace 0,1\rbrace$ \cite{Aaronson2004}. 
Since Clifford gates preserve the Pauli group, 
the action $\ket{\psi} \to {\cal U}\ket{\psi}$ of a Clifford gate ${\cal U}$ 
can be efficiently described by the stabilizers, 
${\cal U} {\cal O}_i {\cal U}^\dagger$ \cite{Note::Heisenberg}, e.g., by  
storing the $\nu_\ell^i, \mu_\ell^i$ in a binary matrix ${\cal M}$ and updating 
their values appropriately~\cite{Aaronson2004}. 

We show that random Clifford circuits can 
elucidate the interplay between transport and entanglement \cite{Znidaric2020}. 
We consider circuits with U$(1)$ symmetry,  
where one time step is defined as the application of $L$ gates 
conserving the total magnetization, $\braket{\psi(t)|\sum_\ell 
Z_\ell|\psi(t)}=\text{const.}$~[Fig.\ 
\ref{Fig::Scheme}]. This property is quite restrictive: While 
the full two-qubit Clifford 
group has $11520$ distinct elements (modulo a global phase), only $64$ conserve 
the total 
Pauli-$Z$ component, see \cite{SuppMat}. 
Due to the U$(1)$ symmetry and the 
Pauli-preserving property of Clifford gates, it turns out that 
transport 
can be understood classically in terms of long-range random 
walks, so called L\'evy flights \cite{Metzler2000, 
Zaburdaev2015, Schuckert2020, Joshi2021}. However, we 
will show that such 
constrained circuits still
generate extensive entanglement, similar to Haar-random 
circuits~\cite{Nahum2017}. 
\begin{figure}[tb]
 \centering
 \includegraphics[width=0.95\columnwidth]{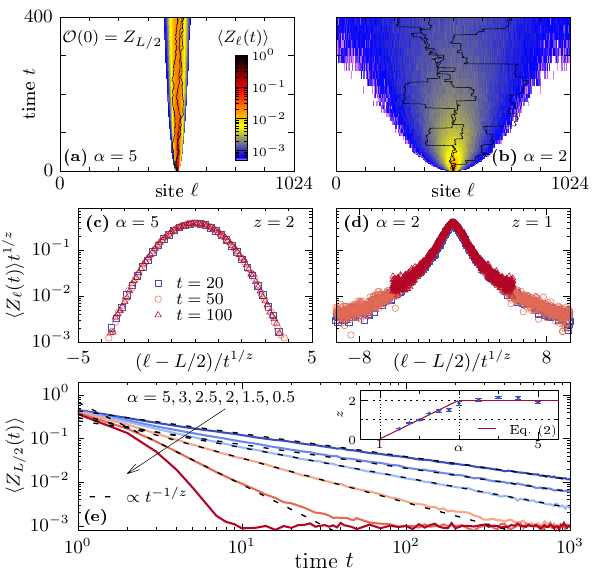}
 \caption{{\bf [(a),(b)]} $\langle Z_\ell(t)\rangle$ 
 averaged over $\sim 10^5$ circuit realizations for $\alpha = 5$ and $\alpha= 
2$ and $L = 1024$. Solid curves indicate individual realizations, i.e., 
random-walks with step-size distribution $\propto 
r^{-\alpha}$. {\bf [(c),(d)]} $\langle Z_\ell(t) \rangle t^{1/z}$ at fixed 
$t$, plotted against $(\ell-L/2)/t^{1/z}$. {\bf (e)} $\langle 
Z_{L/2}(t)\rangle$ 
for different $\alpha$. Dashed lines indicate power law $\propto 
t^{-1/z}$. Inset shows $z$ extracted from the fits and compared to 
Eq.~\eqref{Eq::Levy}.}
 \label{Fig::Transport}
\end{figure}
\begin{figure}[tb]
	\centering
	\includegraphics[width=\columnwidth]{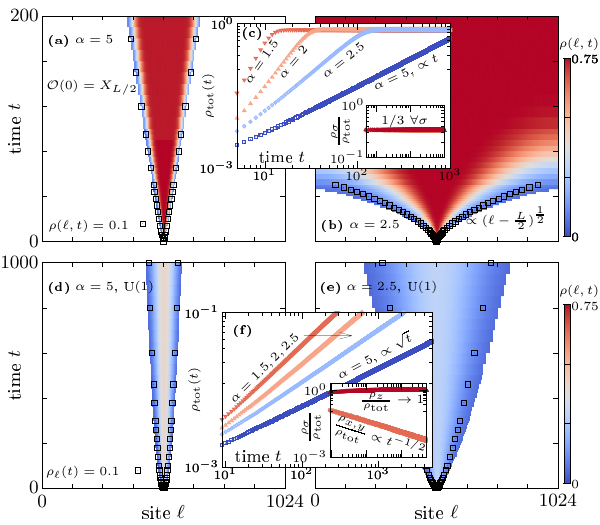}
	\caption{[{\bf (a),(b)}] Averaged $\rho(\ell,t)$ in full Clifford 
		circuits with 
		$\alpha = 5$ and $\alpha = 2.5$, obtained from  
		${\cal O}(0) = X_{L/2}$ with $L = 1024$. Symbols 
		indicate $\rho(\ell,t) = 10^{-1}$. {\bf (c)}
		$\rho_\text{tot}(t)$ for $L = 2048$ and different $\alpha$ (see also  
		\cite{SuppMat}). Inset shows 
		$\rho_\sigma(t)/\rho_\text{tot}(t) \approx 1/3$, i.e., all 
		$\Sigma^\sigma$ contribute equally. [{\bf 
			(d)},{\bf (e)},{\bf (f)}] Analogous data, but for 
U$(1)$-symmetric circuits, where 
		${\cal O}(t)$ spreads significantly slower. This 
		stems from 
		the dominant contribution of $Z$ operators within 
		$\rho_\text{tot}(t)$, cf.\ inset in (f) for 
		$\alpha = 5$. }
	\label{Fig::LC}
\end{figure}

Product states such as $\ket{\rightarrow}^{\otimes L}$ with spins in the 
$x$ direction can be stabilized by operators 
${\cal O}_i = X_i$ ($i = 1,\dots,L$) acting nontrivially only on a 
single site. 
Evolving $\ket{\psi}$ with respect to a random circuit will cause 
the ${\cal O}_i$ to become nonlocal, resulting in increased 
entanglement. Clifford circuits 
are special as they generate 
flat entanglement spectra such that all $S_n$ are equivalent \cite{Fattal2004}. 
While the 
different behaviors of $S_1$ and $S_{n>1}$ demonstrated in \cite{Rakovszky2019} 
therefore cannot be resolved, $S(t)$ 
is nevertheless sensitive to conservation laws and $S(t) \propto \sqrt{t}$ 
was found in Clifford circuits with diffusive 
transport \cite{Znidaric2020}. Here, we show 
that long-range circuits provide an ideal framework to 
study entanglement dynamics also for other transport types. To this end, 
we reiterate the arguments 
to explain the constrained entanglement growth \cite{Rakovszky2019, Huang2020}: 
Consider the 
reduced density matrix $\rho_A$ with $\chi$ nonzero 
eigenvalues $\Lambda_1 \leq \dots \leq \Lambda_\chi$. In the 
presence of hydrodynamic modes with dynamical exponent $z$, 
$\Lambda_\chi$ can be bounded by    
$\Lambda_\chi \gtrsim e^{-\gamma t^{1/z}}$ with some constant $\gamma$, 
where $z=2$ corresponds to diffusion \cite{Rakovszky2019, 
Huang2020}. The bound results from rare contributions to $\ket{\psi(t)}$, 
where a region of length $\xi$ 
around the cut between $A$ and $B$ is in the $\ket{\uparrow}$ state, 
acting as a bottleneck for entanglement as it takes time 
$\propto \xi^z$ for a $\ket{\downarrow}$ to get across the cut. It 
follows that $S_{n \to \infty} = -\log_2 
\Lambda_\chi$ scales as $S_\infty(t) \propto t^{1/z}$ and, due to 
$S_{\infty} \leq S_{n>1} \leq n S_\infty/(n-1)$, all $S_{n>1}(t)$ obey 
this scaling. This is independent of the type of 
time evolution and generalizes to Clifford circuits, 
where $\Lambda_i = \Lambda$ and $S_n(t) \equiv S(t)$.

{\it Hydrodynamics.--}
By varying $\alpha$, 
it is possible to tune the nature of transport. 
Consider a
state $\ket{\psi} = \ket{\rightarrow}^{\otimes L/2-1} 
\ket{\uparrow} \ket{\rightarrow}^{\otimes L/2}$, stabilized by $X_\ell$ for 
$\ell \neq L/2$, and $Z_\ell$ for $\ell = L/2$, cf.\ Fig.\ 
\ref{Fig::Scheme}~(a). The action ${\cal U}{\cal O}_i {\cal 
U}^\dagger$ of U$(1)$-symmetric Clifford gates on the two 
classes of stabilizers is quite different. 
While the $X_\ell$  
becomes nonlocal and 
generates entanglement, the 
stabilizer $Z_{L/2}$ remains of length one 
throughout the circuit \cite{SuppMat}. Specifically, the $Z$ 
operator performs 
$\alpha$-dependent random walks, i.e., L\'evy flights 
\cite{Schuckert2020, Joshi2021}, 
examples of which are shown in Figs.\ \ref{Fig::Transport}~(a), (b) for $\alpha 
= 5$ and $\alpha = 2$. Consequently, at a given time, 
there will be a site $\ell$ 
with $\braket{\psi(t)|Z_\ell|\psi(t)} = 1$ unentangled with the rest of the 
system \cite{NoteZZ}.

Simulating 1$d$ circuits with $L = 1024$, Figs.\ 
\ref{Fig::Transport}~(a), (b) show the circuit-averaged 
value $\langle Z_\ell(t) \rangle$ for $\sim10^5$ random 
realizations of ${\cal U}Z_{L/2}{\cal U}^\dagger$, highlighting 
a change from local to non-local when reducing $\alpha$. 
Analyzing 
$\langle 
Z_\ell(t)\rangle$ at fixed $t$, we find Gaussian profiles for 
$\alpha = 5$ that collapse when 
rescaled appropriately [Fig.\ \ref{Fig::Transport}~(c)], indicating 
diffusion. In contrast, $\langle Z_\ell(t)\rangle$ 
is 
non-Gaussian for $\alpha = 2$ but rather described by a Lorentzian, signaling 
superdiffusive transport \cite{Kloss2019, 
Schuckert2020}. (See 
\cite{SuppMat} 
for other $\alpha$ and 2$d$ circuits.) 
The $\alpha$-dependent transport regimes are also reflected in the 
decay at $\ell = L/2$,  
$\langle Z_{L/2}(t)\rangle \propto 
t^{-1/z}$, 
where $z$ approximately follows the L\'evy-flight prediction 
\cite{Schuckert2020, Joshi2021, NoteAlpha}, 
\begin{equation}\label{Eq::Levy}
  z = \begin{cases}
                                      2, \ &\alpha \geq 3; \\
                                      \alpha-1, \ &1 < \alpha \leq 3, 
                                    \end{cases} 
\end{equation}
with no hydrodynamic tail for $\alpha \leq 1$ [Fig.~\ref{Fig::Transport}~(e)].
Clifford and U$(1)$-symmetric 
Haar-random gates are expected to yield the same circuit-averaged
$\langle Z_\ell(t)\rangle$. In contrast, individual circuit realizations 
differ since Haar gates distribute the $Z$ excitation 
smoothly over multiple sites whereas Clifford gates yield sharp random walks. 
The transport behavior in Fig.\ \ref{Fig::Transport} agrees 
qualitatively with the emergent quantum hydrodynamics observed in long-range 
Hamiltonian systems \cite{Schuckert2020, Joshi2021}. Even 
though transport in the Clifford case is a purely classical process, the 
average coarse-grained type of hydrodynamics, both in the circuit and the 
Hamiltonian model, is especially at high temperatures mainly set by the range 
of the interactions (i.e., by 
$\alpha$), and not so much by the microscopic dynamics.

{\it Operator spreading.--}
While ${\cal U} Z_\ell {\cal U}^\dagger$ remains a single-site operator for 
U$(1)$-symmetric Clifford gates [Fig.\ \ref{Fig::Transport}], we now consider 
${\cal O} = X_\ell$. 
Generally, ${\cal 
O}(t) =\sum_{\cal 
S} \alpha_{\cal S}(t) {\cal S}$ can be written in the basis 
of the $4^L$ Pauli strings ${\cal S}$. Evolution under Haar-random 
gates increases the number of nonzero $\alpha_{\cal 
S}(t)$  
\cite{Nahum2017,Nahum2018,Keyserlingk2018}, leading to
operator entanglement \cite{Zanardi2001}. 
In contrast, Clifford 
gates
map Pauli operators to each other, ${\cal O}(t) = 
\delta_{{\cal S},{\cal O}(t)} {\cal S}$, with no operator entanglement. 
However, ${\cal O}(t)$ will 
become nonlocal, manifested by its growing support 
$\rho_\text{tot}(t) =  \frac{1}{L} \sum_{\ell,\sigma} 
\rho_{\sigma}(\ell,t)$, where 
$\rho_{\sigma}(\ell,t) =  
\text{tr}[{\cal 
O}_\ell(t)\Sigma^\sigma]/2$, 
${\cal O}_\ell(t)$ is the matrix at 
position $\ell$ 
in the string, and 
$\Sigma^{\sigma} = \lbrace
X, Y, 
Z\rbrace$, $\sigma = x,y,z$. 

\begin{figure}[t] 
    \centering
     \includegraphics[width=0.95\columnwidth]{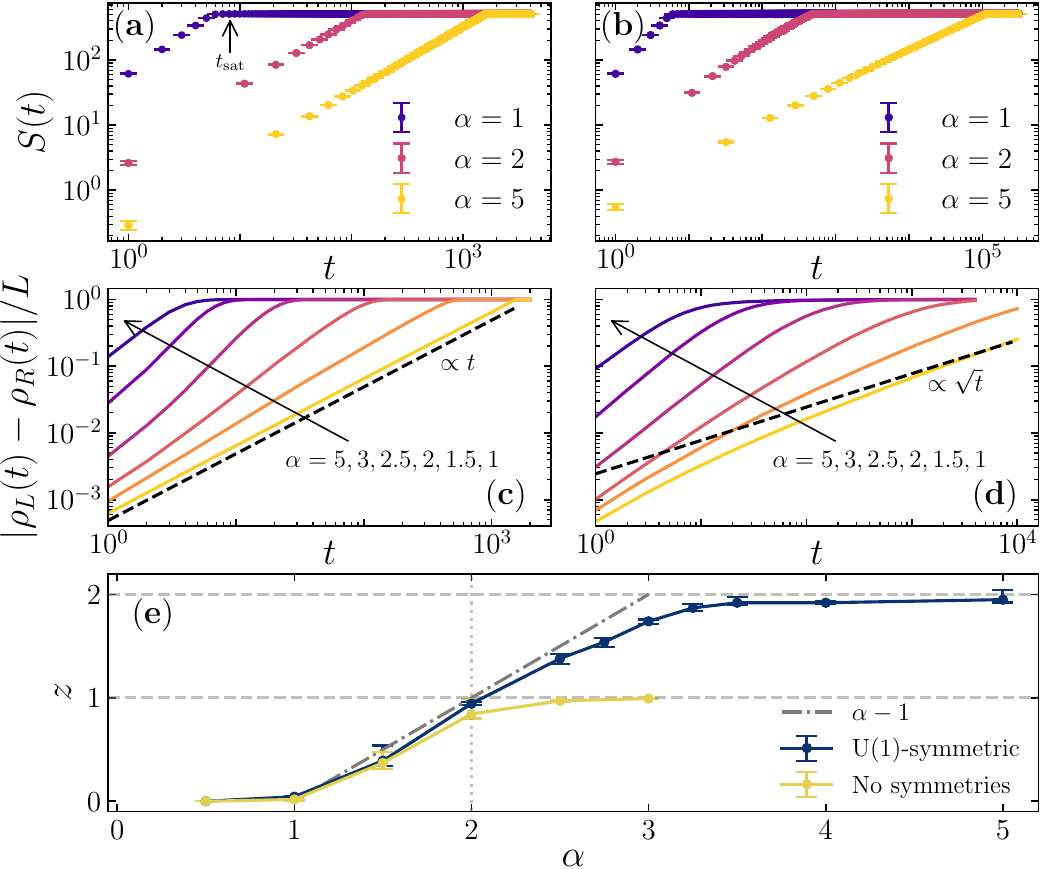}
     \caption{[\textbf{(a),(b)}] $S(t)$ for different $\alpha$ in asymmetric and
     U$(1)$-symmetric circuits with 
     $L=1024$ and open boundaries. For $\alpha\geq3$, 
     we expect short-range behavior: 
     $z=2$ with U$(1)$ symmetry and $z=1$ without.
     [\textbf{(c),(d)}] Normalized difference between 
left (right) endpoints of ${\cal O}(t)$ for $L=2048$. The 
dashed curves indicate $\propto t$ ($\propto \sqrt{t}$) scaling. 
     \textbf{(e)} $z$ versus $\alpha$ for circuits with and 
without conservation law. 
     The deviations from \cref{Eq::Levy} around $\alpha=3$ may be 
due to logarithmic corrections to transport~\cite{Schuckert2020}.
}
     \label{fig:entanglement}
\end{figure}

Considering ${\cal O}(0) = X_{L/2}$, we plot $\rho(\ell,t) = \sum_{\sigma} 
\rho_\sigma(\ell,t)$ in Fig.\ \ref{Fig::LC}, which is a measure 
for the out-of-time-ordered correlator between operators 
at sites $\ell$ and $L/2$ \cite{Xu2022}.
For circuits without conservation law [Figs.\ 
\ref{Fig::LC}~(a),(b)], we observe a linear light cone for $\alpha = 5$, 
while a power-law light cone emerges for $\alpha = 2.5$, in agreement with 
the phase diagram in \cite{Zhou2020}. 
Correspondingly, we find $\rho_\text{tot}(t)\propto t$ at $\alpha = 5$ and 
faster growth for smaller $\alpha$ [Fig.\ 
\ref{Fig::LC}~(c)], see also \cite{SuppMat}.
In the bulk of the light cone, we observe full scrambling
with 
$\rho(\ell,t) \to 3/4$ and $\rho_\sigma(t)/\rho_\text{tot}(t) 
\approx 1/3$ [insets in Fig.\ \ref{Fig::LC}~(c)], where $\rho_\sigma(t) = 
\sum_\ell 
\rho_\sigma(\ell,t)$ is the Pauli-component resolved support. 
Speaking differently, the interior of the light cone has 
reached an equilibrium distribution where local $X,Y,Z$ operators are equally 
likely. We will discuss the dynamics of the light-cone edges 
further below in the context of Fig.\ \ref{fig:entanglement}.

Next, turning to U$(1)$-symmetric gates, the behavior of 
$\rho(\ell,t)$ changes drastically [Figs.\ \ref{Fig::LC}~(d),(e)]. 
Namely, operator spreading is significantly slower 
and resembles the transport behavior of the 
conserved quantity [Eq.\ \eqref{Eq::Levy}], with a 
diffusive (superdiffusive) light cone for 
$\alpha = 5$ ($\alpha = 2.5$), also reflected in the growth of 
$\rho_\text{tot}(t)$ [Fig.\ \ref{Fig::LC}~(f)]. 
This is due to the properties of the 
U$(1)$-symmetric Clifford gates, which cause ${\cal 
O}(t)$ to be dominated by $Z$ operators. 
Given the initial operator ${\cal 
O}(0) = X_{L/2}$ with a single $X$ at $\ell = L/2$ and identity 
operators on all other sites, it is in fact significantly 
more likely that a random gate will 
generate more $Z$ than $X$, $Y$ 
operators and thereby increase the overall 
share of $Z$ in ${\cal O}(t)$, see 
\cite{SuppMat} for details. This 
is shown in the inset of Fig.\ \ref{Fig::LC}~(f) for $\alpha = 
5$, where we find $\rho_z(t) 
\propto t^{1/2}$ while $\rho_{x,y}(t) = \text{const.}$ such that 
$\rho_z(t)/\rho_\text{tot}(t) \to 1$. The inhomogeneous composition of ${\cal 
O}(t)$ differs from the unsymmetric case where $X$, $Y$, $Z$ occur with 
equal probability [Fig.\ \ref{Fig::LC}~(c)]. The large fraction of $Z$ 
operators behaves similarly to Fig.\ \ref{Fig::Transport}, 
leading to narrower light cones compared to circuits without conservation law. 
Furthmore, studying the bulk of the light cone, we find 
that $\rho(\ell,t) < 3/4$ in the U$(1)$-symmetric case [Fig.\ 
\ref{Fig::LC}~(d),(e)]. This indicates that at least on the time scales shown 
here, the operator string is not fully scrambled and contains on average more 
identity operators than in the case without conservation law.

The operator spreading in 
U$(1)$-symmetric Clifford circuits is notably simpler 
compared to the Haar-random case, where the conserved charges 
lag behind the light-cone front which propagates 
quickly due to nonconserved 
operators \cite{Khemani2018}. 
While Clifford gates fail to capture this aspect of generic quantum 
dynamics, the 
simplified description is helpful to 
understand the constrained entanglement dynamics since the light cones in Fig.\ 
\ref{Fig::LC} upper bound the growth of $S(t)$~\cite{Nahum2017}.

{\it Entanglement dynamics.--}
Choosing 
$\ket{\psi(0)} = \ket{\rightarrow}^{\otimes L}$, we study $S(t) =  
\text{rank}({\cal M}_{L/2}) - L/2$ for a 
half-system cut, where
${\cal M}_{L/2}$ denotes the stabilizer matrix of the first $L/2$ 
sites \cite{Nahum2017, Hamma2005,Note_Mod2}. From this expression, it is 
clear that $S(t)$ depends on the collective dynamics of $\ket{\psi(t)}$'s 
stabilizers.
Since $\ket{\psi(0)}$ is a superposition of 
all symmetry sectors, $S(t\to \infty) \approx L/2$ saturates 
at the same value in circuits with and without the conservation law [Figs.\ 
\ref{fig:entanglement}~(a),(b)]. We find it 
convenient to analyze 
the $\alpha$-dependence of $S(t)$ by extracting the saturation time 
$t_\text{sat} \propto L^z$ for different $L$, implying an asymptotic 
scaling $S(t) \propto t^{1/z}$.
The obtained values of $z$ are 
summarized in \cref{fig:entanglement}~(e). In the case of 
U$(1)$-symmetric circuits, we find that 
the transport behavior is 
reflected in the entanglement dynamics and 
$z$ is reasonably well 
described by Eq.\ \eqref{Eq::Levy}. 
In addition, while we recover $z \to 1$ in unsymmetric 
circuits for $\alpha \geq 3$, as expected for short-range models 
\cite{Nahum2017}, the scaling behaviors of circuits with 
and without conservation law become similar for $\alpha \lesssim 2$, with 
all discrepancies in $z$ estimates contained within error bars.

At small $\alpha$, transport is fast enough that entanglement 
growth is mainly dictated by the gate range and not by the conservation law.
Specifically, at $\alpha = 2$ we have $z \approx 1$ and the bound 
$\Lambda_{\chi} \gtrsim 
e^{-\gamma t^{1/z}}$ due to transport becomes comparable to the 
typical value $\sim e^{-\gamma t}$ expected given the ballistic $S_{1}(t)$ in 
generic circuits~\cite{Rakovszky2019}. The behavior of the edges of 
the light cone can provide further 
quantitative insights. Specifically, we study the endpoints $\rho_{L(R)}(t)$ of 
a string 
${\cal 
O}(t)$, i.e., the left(right)most $\ell$ where ${\cal O}_\ell(t)$ is 
nonidentity. Once a nontrivial part of ${\cal 
O}(t)$ extends across the cut, entanglement may in principle increase. One 
might 
therefore expect that 
$\rho_{L(R)}(t)$ is more relevant for $S(t)$ than 
$\rho_\text{tot}(t)$ [Fig.\ \ref{Fig::LC}]. As shown in Figs.\ 
\ref{fig:entanglement}~(c) and (d), we find that $|\rho_L(t)-\rho_R(t)|/L$ 
behaves very differently in symmetric and unsymmetric circuits for $\alpha = 
5$, but grows with 
roughly comparable rate if $\alpha$ is small (see also \cite{SuppMat}), 
which is consistent with the observed similar growth rate of 
entanglement.

We expect the relation between transport and entanglement to 
carry over to R\'{e}nyi entropies 
$S_{n>1}(t)$ in generic systems with a conserved quantity, see 
\cite{SuppMat} for some evidence in a long-range tilted field 
Ising model.
Since Clifford gates form unitary 
3-designs~\cite{Harrow2009,Webb2015}, they give
the same ``annealed'' R\'{e}nyi-2 entropy 
$S_{2}^{(a)} = -\log \overline{\mathrm{tr}_{A}
\rho_{A}^{2}}$ as a Haar-random circuit. Although $S_{2}^{(a)} \leq 
\overline{S_{2}}$ only
lower bounds the average $\overline{S_{2}}$, in
U$(1)$-symmetric Haar-random circuits it displays the
same $\sqrt{t}$ growth as $\overline{S_{2}}$ \cite{Rakovszky2019}, consistent 
with small sample-to-sample fluctuations of 
$S_{2}(t)$.

Let us comment on the deviations in 
\cref{fig:entanglement}~(e) from
the prediction \eqref{Eq::Levy}, most 
pronounced near $\alpha = 3$. Even for $L \sim 
10^3$ presented here, we 
observe a drift of $z$ with $L$. We attempt 
to account for these finite-size effects 
by restricting the data 
to $L \leq L_{\mathrm{end}}$ and 
extrapolating $z(L_{\mathrm{end}})$ 
to $1/L_{\mathrm{end}} \to 0$. For details, including how we
obtain the error bars, see \cite{SuppMat}.
Precisely at $\alpha = 3$, transport can 
receive logarithmic 
corrections~\cite{Schuckert2020}, which may also explain 
the faster entanglement growth. 
Repeating our analysis, but with 
$S(t) \sim t^{1/z} \sqrt{\log{t}}$, we obtain $z = 1.91(1)$ much 
closer to $z=2$~\cite{NoteLogCorrection}. The 
marginality at $\alpha \approx 3$ is also 
reflected in the development of non-Gaussian 
tails in both $\langle Z_\ell(t) \rangle$ 
and $\rho(\ell,t)$~\cite{SuppMat}.

{\it Conclusion.--}
We have studied the interplay of transport and 
entanglement dynamics in long-range random Clifford circuits with U$(1)$ 
symmetry. We demonstrated that the emerging 
transport regimes with dynamical exponent $z$ reflect themselves in the growth 
of entanglement as $S(t) 
\propto t^{1/z}$, generalizing earlier work that has focused on diffusive 
systems with $z = 2$ \cite{Rakovszky2019}. 
While we expect this result to hold  
also in more generic 
Haar-random circuits or chaotic quantum systems for $S_{n>1}(t)$, we here 
provided a simplified picture specific to the Clifford framework, where 
operator strings become dominated 
by the conserved quantity leading to narrower light cones.
While transport in Clifford circuits turned out to be purely 
classical, their efficient simulability may 
suggest the study of possible connections with recent state-of-the-art methods 
to capture transport coefficients \cite{Rakovszky2022, White2018, Wurtz2018, 
Ye2020, Leviatan2017, Kvorning2022, Keyserlingk2022}, and to 
better understand the role of entanglement and the differences to full 
thermalizing quantum dynamics~\cite{Farshi2022}.

A promising research direction is to consider 
entanglement dynamics in Clifford circuits with other gate 
sets or conservation laws, potentially giving rise to 
localization \cite{Chandran2015}, as well as adding measurements which can 
induce nonequilibrium phases in 
circuits with symmetry \cite{Agrawal2021, Bao2021}. Studying 
the impact of 
sporadic non-Clifford gates, acting as seeds of chaos \cite{Zhou2020_3}, is 
another natural avenue. 
Finally, it would 
be interesting if 
the transport-dependent entanglement growth is observable 
in quantum-simulator experiments, where diffusion and superdiffusion 
can be realized 
\cite{Joshi2021, Wei2021} and the R\'enyi-$2$ entropy is accessible for small 
systems \cite{Islam2015, Linke2018}. 

\begin{acknowledgments}
 {\it Acknowledgments.--} 
We sincerely thank Lluis Masanes for a helpful comment.
This work was funded by the 
European Research 
Council 
(ERC) under the European Union's Horizon 2020 research and innovation programme
(Grant agreement No.\ 853368). J.R.\ also received funding 
from the European
Union's Horizon Europe programme
under the Marie Sk\l odowska-Curie grant agreement
No. 101060162. O.L.\ acknowledges support by UK Research 
and 
Innovation (UKRI) [grant number MR/T040947/1].
\end{acknowledgments}

\clearpage

\onecolumngrid

% %%%%%%%%%%%%%%%%%%%%%%%%%
% % Supplemental Material %
% %%%%%%%%%%%%%%%%%%%%%%%%%
 
\setcounter{figure}{0}
\setcounter{equation}{0}
 \renewcommand*{\citenumfont}[1]{S#1}
\renewcommand*{\bibnumfmt}[1]{[S#1]}
\renewcommand{\thefigure}{S\arabic{figure}}
\renewcommand{\theequation}{S\arabic{equation}}

\section*{Supplemental material}

\subsection{Structure of the Clifford group}

Let us provide additional explanations on random Clifford circuits. To begin 
with, we note that the Pauli group ${\cal P}_L$ on $L$ qubits is 
generated by $L$-fold tensor products 
of the Pauli matrices, 
\begin{equation}\label{Eq::Pauli_Mats}
 I = \begin{pmatrix}
      1 & 0 \\
      0 & 1 
     \end{pmatrix}\ ,\quad 
    X = \begin{pmatrix}
         0 & 1 \\
         1 & 0 
        \end{pmatrix}\ ,\quad  
    Y = \begin{pmatrix}
         0 & -i \\
         i & 0 
        \end{pmatrix}\ ,\quad  
    Z = \begin{pmatrix}
         1 & 0 \\
         0 & -1 
        \end{pmatrix}\ . 
\end{equation}
The Clifford group ${\cal C}_L$ on $L$ qubits is then defined as the group that 
preserves the Pauli group ${\cal P}_L$ under conjugation, quotiented by U$(1)$ 
to account for a global phase.

\subsubsection{Clifford group on 2 qubits}

The two-qubit Clifford group ${\cal C}_2$ can be generated by the gates 
$\lbrace P, H, 
\text{CNOT}_{0,1} \rbrace$, where 
\begin{equation}
P = \begin{pmatrix}
     1 & 0 \\
     0 & i 
    \end{pmatrix}\ ,\quad
H = \begin{pmatrix}
      1 & 1 \\
      1 & -1
     \end{pmatrix}\ ,\quad
     \text{CNOT}_{0,1} = \begin{pmatrix}
                          1 & 0 & 0 & 0 \\
                          0 & 1 & 0 & 0 \\
                          0 & 0 & 0 & 1 \\
                          0 & 0 & 1 & 0 
                         \end{pmatrix}\ , \quad
             \text{CNOT}_{1,0} = \begin{pmatrix}
                          1 & 0 & 0 & 0 \\
                          0 & 0 & 0 & 1 \\
                          0 & 0 & 1 & 0 \\
                          0 & 1 & 0 & 0 
                         \end{pmatrix}\ . 
\end{equation}
For convenience we further define the composite gates $W$ and $V$, 
\begin{equation}
 W = H \cdot P\ ,\qquad V = W \cdot W = H \cdot P \cdot H \cdot P\ .   
\end{equation}
The two-qubit Clifford group can then be structured into different classes, 
characterized by their number of two-qubit gates \cite{Corcoles2013S}. The 
first 
class contains solely single-qubit gates,  
\begin{equation}\label{Eq::Cliff1}
 (h_0 \otimes h_1)(v_0\otimes v_1)(p_0\otimes p_1)\ ,\ \text{with}\ h_i 
 \in \{I,H\},\ v_i \in \{I,V,W\}\ \text{and}\ p_i \in \{I,X,Y,Z\}\ , 
\end{equation}
which results in $2^2\times 3^2 \times 4^2 = 24^2 = 576$ distinct gates. 
The second class requires one CNOT gate, 
\begin{equation}
 (h_0\otimes h_1) (v_0\otimes v_1) \text{CNOT}_{0,1} (v'_0\otimes v'_1) 
 (p_0\otimes p_1)\ ,\ \text{with}\ h_i \in \{I,H\}\ ,\ v_i, v'_i \in \{I,V,W\}\ 
,\ p_i \in \{I,X,Y,Z\}\ , 
\end{equation}
which contains $2^2 \times 3^2 \times 3^2 \times 4^2 = 5184$ gates.
The third class comprises sequences with two CNOT gates,  
\begin{equation}
 (h_0\otimes h_1) (v_0\otimes v_1) \text{CNOT}_{0,1}\text{CNOT}_{1,0} 
 (v'_0\otimes v'_1) (p_0\otimes p_1)\ ,\ \text{with}\ h_i \in \{I,H\},\ v_i, 
v'_i \in \{I,V,W\}\ ,\ p_i \in \{I,X,Y,Z\}\ , 
\end{equation}
which yields $2^2 \times 3^2 \times 3^2 \times 4^2 = 5184$ gates. 
Eventually, the fourth class requires three CNOT gates, 
\begin{equation}\label{Eq::Cliff4}
 (h_0\otimes h_1) (v_0 \otimes v_1) \text{CNOT}_{0,1} \text{CNOT}_{1,0} 
\text{CNOT}_{0,1} (p_0\otimes p_1)\ ,\ \text{with}\ h_i \in \{I,H\}, v_i, \in 
\{I,V,W\}, p_i \in \{I,X,Y,Z\}\ , 
\end{equation}
which contains $2^2 \times 3^2 \times 4^2 = 576$ gates. 
In total, there are thus $11520$ distinct 2-qubit Clifford gates.   

In practice, these $11520$ distinct gates can be stored in a look-up table.
Every application of a two-qubit Clifford gate in the circuit then corresponds
to selecting and carrying out a random element of the look-up table.
Alternatively, another useful approach to randomly select an element of ${\cal
C}_2$ has been presented in \cite{Koenig2014S}. In essence, it consists of
generating a suitable symplectic matrix, which upon multiplication with the
stabilizer tableau, implements the action of a random gate. This approach is
particularly beneficial if one is interested in Clifford gates on more than two
qubits since $|{\cal C}_{n>2}|$ is too large to be stored in a look-up table.
In this paper,  we use both approaches complementarily.  

\subsubsection{2-qubit Clifford gates that conserve $\bra{\psi(t)}Z_1 + 
Z_2\ket{\psi(t)}$}

In this paper, we are mainly interested in the interplay between transport and 
entanglement growth. To this end, we consider circuits with U$(1)$ symmetry 
that conserve the total Pauli-$Z$ component, such that magnetization exhibits 
hydrodynamic transport. Given the decomposition of the full two-qubit Clifford 
group in Eqs.\ \eqref{Eq::Cliff1} - \eqref{Eq::Cliff4}, the Clifford gates 
that conserve magnetization can be written as follows. 
The first class consists of single-qubit gates, 
\begin{equation}\label{Eq::Cliff12}
 (h_0 \otimes h_1)(p_0\otimes p_1)\ ,\ \text{with}\ h_i \in \{I,P\},\ 
 \text{and}\ p_i \in \{I,Z\}\ , 
\end{equation}
and contains $2^2\times 2^2 = 4^2 = 16$ distinct gates. The second class 
requires one $\text{CNOT}$ gates,
\begin{equation}
 (h_0\otimes h_1) \text{CNOT}_{0,1} (I \otimes W) (p_0\otimes p_1)\ ,\ 
 \text{with}\ h_0 \in \{I,P\}\ ,\ h_1, \in \{V,H\}\ ,\ p_i \in \{I,Z\}\ , 
\end{equation}
and contains $2 \times 2 \times 2^2 = 16$ distinct gates.
The third class requires two \text{CNOT} gates, 
\begin{equation}
 (h_0\otimes h_1)  \text{CNOT}_{0,1}\text{CNOT}_{1,0} (I\otimes W) 
 (p_0\otimes p_1)\ ,\ \text{with}\ h_0 \in \{H,V\}\ ,\ h_1 \in \{I,P\},\ p_i 
\in \{I,Z\}\ , 
\end{equation}
and contains $2 \times 2 \times 2^2 = 16$ gates.
Finally, the fourth class requires three \text{CNOT} gates, 
\begin{equation}\label{Eq::Cliff12_2}
 (h_0\otimes h_1) \text{CNOT}_{0,1} \text{CNOT}_{1,0} \text{CNOT}_{0,1} 
 (p_0\otimes p_1)\ ,\ \text{with}\ h_i \in \{I,P\}\ ,\ p_i \in \{I,Z\}\ , 
\end{equation}
and contains $2^2 \times 2^2 = 16$ gates.
Thus, there are $64$ distinct 2-qubit Clifford gates which conserve 
the magnetization $\bra{\psi(t)}Z_1 + Z_2\ket{\psi(t)}$. We note that this is 
distinctly smaller than 
the size of the full two-qubit Clifford group, $64 \ll |{\cal C}_{2}| = 
11520$. Moreover, regarding the production of entanglement, let us note that 
in the full two-qubit Clifford group only $576$ gates are separable [Eq.\ 
\eqref{Eq::Cliff1}], which corresponds to a fraction of $576/11520 = 
0.05$. In contrast, in the case of gates that conserve magnetization, $16/64 = 
0.25$ gates are separable [Eq.\ \eqref{Eq::Cliff12}].  
Thus, if one considers Clifford circuits with U$(1)$ symmetry, the application
of a random gate will, on average, produce less entanglement compared to
Clifford circuits without conservation law. This has the effect that, even in
regimes where the dynamical critical exponent is the same for symmetric and
asymmetric circuits, symmetric circuits will typically take longer (by some
$\mathcal{O}(1)$ factor) to reach the steady-state value than asymmetric 
circuits.

It is also instructive to study the action of the 
U$(1)$-symmetric Clifford gates in Eqs.\ \eqref{Eq::Cliff12} - 
\eqref{Eq::Cliff12_2} on Pauli operators, ${\cal U} ({\cal O}_1 \otimes {\cal 
O}_2){\cal U}^\dagger$. Given two lattice sites, as well as the Pauli and 
identity operators in Eq.\ \eqref{Eq::Pauli_Mats}, there are $2^4$ different 
configurations to consider. First of all, it is obvious that with probability 
$p 
= 1$,
\begin{equation}
I \otimes I \longrightarrow I 
\otimes I\ (p = 1)\ , 
\end{equation}
i.e., given identity operators on both lattice sites, this 
configuration remains unchanged for all $64$ possible U$(1)$-symmetric Clifford 
gates. This result naturally holds for Clifford gates without conservation 
law as well. Crucially, for other nontrivial initial operator configurations, 
the effect of the U$(1)$ conservation law becomes apparent. In particular, we 
have,
\begin{equation}\label{Eq::ZZ}
\begin{rcases}
Z \otimes \mathbb{1} \\
\mathbb{1}
\otimes Z
 \end{rcases}
\longrightarrow \begin{cases}
                                              Z
\otimes \mathbb{1}\ (p = 1/2)  \\
\mathbb{1}
\otimes Z\ (p = 1/2)
\end{cases}\ ,\quad Z \otimes Z \longrightarrow Z 
\otimes Z\ (p = 1)\ , 
\end{equation}
which highlights the fact that a single $Z$ operator can 
perform jumps between different lattice sites, but no other operators are 
created in the process. As a consequence, if one starts with an isolated $Z$ 
operator, the application of U$(1)$-symmetric Clifford gates will lead to a 
random-walk of the $Z$ operator, but the operator string will remain of length 
one throughout the entire circuits.

In contrast, if the initial configuration 
contains solely $X$ or $Y$ operators, the U$(1)$-symmetric Clifford gates 
cannot 
produce new $Z$ operators,
\begin{equation}
 \begin{rcases}
  X \otimes X \\
  Y \otimes Y \\
  X \otimes Y \\
  Y \otimes X
 \end{rcases}\longrightarrow
 \begin{cases}
  X \otimes X\ &(p = 1/4)\\
  Y \otimes Y\ &(p = 1/4) \\
  X \otimes Y\ &(p = 1/4) \\
  Y \otimes X\ &(p = 1/4)
 \end{cases}\ . 
\end{equation}
Finally, the remaining 8 configurations transform according 
to, 
\begin{equation}\label{Eq::Last}
 \begin{rcases}
  X \otimes \mathbb{1} \\
  \mathbb{1} \otimes X \\
  Y \otimes \mathbb{1} \\
  \mathbb{1} \otimes Y \\
  X \otimes Z \\
  Z \otimes X \\
  Y \otimes Z \\
  Z \otimes Y 
 \end{rcases}\longrightarrow
 \begin{cases}
 X \otimes \mathbb{1}\ &(p = 1/8)\\
  \mathbb{1} \otimes X\ &(p = 1/8)\\
  Y \otimes \mathbb{1}\ &(p = 1/8) \\
  \mathbb{1} \otimes Y\ &(p = 1/8) \\
  X \otimes Z\ &(p = 1/8) \\
  Z \otimes X\ &(p = 1/8) \\
  Y \otimes Z\ &(p = 1/8) \\
  Z \otimes Y\ &(p = 1/8) 
 \end{cases}\ . 
\end{equation}
These update rules for two-site Pauli operators under the 
action of U$(1)$-symmetric Clifford gates also allows an understanding of our 
finding in the context of Figs.\ 3~(d)-(f). In particular, starting 
with an isolated $X$ operator, with probability $p = 4 \times 1/8 = 1/2$, a $Z$ 
operator is created by the first Clifford gate acting on $X \otimes I$. Since 
this $Z$ operator remains conserved when a gate acts on $Z \otimes I$ [cf.\ 
Eq.\ \eqref{Eq::ZZ}], and since the initial operator string contains many $I$ 
operators, it is highly probable that more $Z$ operators are created due to 
Eq.\ \eqref{Eq::Last} and these $Z$ operators then dominate the spreading of 
the light cone.

\subsection{Additional data on hydrodynamics and operator spreading in 
long-range 1$d$ Clifford circuits}
\begin{figure}[tb]
 \centering
 \includegraphics[width=0.85\textwidth]{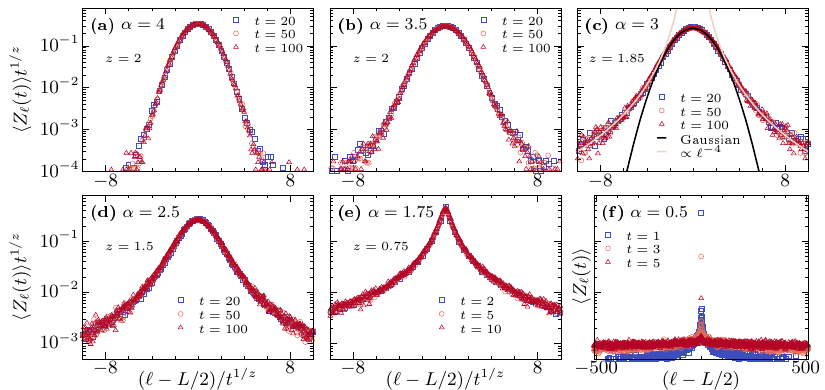}
 \caption{[{\bf (a)-(e)}] Rescaled density profile $\langle Z_\ell(t)\rangle 
t^{1/z}$
versus $(\ell - L/2)/t^{1/z}$ in long-range circuits with different values 
 of $\alpha$. For $\alpha = 3$, we observe that while the bulk is well 
described by a Gaussian, the 
distribution exhibits heavy tails $\propto \ell^{-4}$. {\bf (f)} $\langle 
Z_\ell(t)\rangle$ versus $\ell-L/2$ in circuits with $\alpha = 0.5$. The system 
thermalizes in a few time steps, resulting in a flat distribution.  }
 \label{Fig_Hydro}
\end{figure}

\subsubsection{Transport properties}

In addition to the data presented in Fig.\ 2 in the main 
text, we provide further numerical results on transport in long-range Clifford 
circuits 
in Fig.\ \ref{Fig_Hydro}. We emphasize that this data can be understood 
as resulting from the time evolution of an initial state of the form 
$\ket{\psi(0)} = \ket{\rightarrow}^{\otimes L/2-1} 
\ket{\uparrow} \ket{\rightarrow}^{\otimes L/2}$.  
However, due to the particular nature of U$(1)$-symmetric Clifford gates, i.e., 
the fact that the stabilizer ${\cal O}_i = Z_{L/2}$ will remain of length one 
throughout the entire circuit, it is in fact not necessary to 
study $\ket{\psi(t)}$, but just to keep track of the random walk ${\cal 
U}Z_{\ell}{\cal U}^\dagger$ of the isolated $Z$ operator. Averaging over many 
random 
circuit realizations yields the expectation value $\langle Z_\ell(t)\rangle$.

In Fig.\ 
\ref{Fig_Hydro}~(a)-(e), the rescaled expectation value $\langle 
Z_\ell(t)\rangle t^{1/z}$ is plotted versus $(\ell-L/2)/t^{1/2}$ for fixed 
times $t$ and different values of $\alpha$. Generally, we find a convincing 
agreement with the theoretical L\'evy-flight prediction [Eq.\ 
(2) in main text]. In particular, we observe approximate 
Gaussian profiles for $\alpha > 3$ that collapse for the diffusive value $z = 
2$. Moreover, for $1 < \alpha < 3$, the profile becomes non-Gaussian with a 
pronounced peak at $\ell = L/2$ and collapses for $z = \alpha -1$. 

Let us now comment on $\alpha = 3$ [Fig.\ \ref{Fig_Hydro}~(c)].  
In contrast to the prediction of $z = 2$, we find 
that the numerically obtained 
value $z \approx 1.85$ yields a much more 
convincing data collapse. These discrepancies might be due to finite-size 
and finite-time effects, which we expect to be most pronounced at the phase 
boundary. Furthermore, while the bulk of $\langle Z_\ell(t)\rangle$ is 
well described by a Gaussian, consistent with the emergence of 
diffusive transport for $\alpha \geq 3$, we observe that $\langle 
Z_\ell(t)\rangle$ exhibits heavy non-Gaussian tails 
decaying as $\propto 
\ell^{-4}$, which are well-known to occur 
for L\'evy flights \cite{Schuckert2020S}. It would be interesting to 
understand in more detail the 
potential impact of such tails on the dynamics of entanglement $S(t)$ 
discussed in Fig.\ 4. In 
particular, as shown in Fig.\ \ref{Fig_Rho}, the operator 
spreading quantified by $\rho(\ell,t)$ likewise develops such non-Gaussian 
tails. Let us 
note that for larger values of $\alpha$, such as $\alpha = 5$ considered in 
Fig.\ 2~(c) in the main text, the density distribution is 
well described by a Gaussian without heavy tails (at least within the 
limitations set by the statistical fluctuations).

Eventually, in Fig.\ \ref{Fig_Hydro}~(f), we show $\langle Z_\ell(t)\rangle$  
obtained in highly non-local circuits with $\alpha = 0.5$. Consistent with the 
absence of a hydrodynamic tail of $\langle Z_{L/2}(t)\rangle$, cf.\ Fig.\ 
2~(e), we find that the single $Z$ excitation spreads over 
the entire 
system within a few time steps, resulting in a flat distribution.

\subsubsection{Operator spreading}

\begin{figure}[tb]
 \centering
 \includegraphics[width=0.7\textwidth]{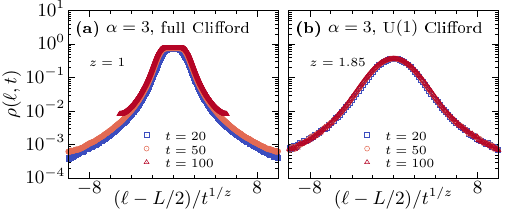}
 \caption{Circuit-averaged $\rho(\ell,t)$ versus rescaled variable 
$(\ell-L/2)/t^{1/z}$, obtained analogous to Fig.\ 3 of the main 
text. Data is shown for circuits with $L = 1024$ and $\alpha = 3$ for {\bf (a)} 
Clifford gates without 
conservation law and 
{\bf (b)} U$(1)$-symmetric gates.}
 \label{Fig_Rho}
\end{figure}

To proceed, we also provide additional data on the operator spreading ${\cal 
U}X_{L/2}{\cal U}^\dagger$ of an initially isolated $X$ operator, analogous to 
Fig.\ 3 of the main text. In Fig.\ \ref{Fig_Rho}~(a) and (b), we 
focus on $\alpha = 3$ and show cuts of $\rho(\ell,t)$ (see definition in main 
text) at fixed times for Clifford circuits without conservation law as well as 
circuits with U$(1)$-symmetric gates. In the former case [Fig.\ 
\ref{Fig_Rho}~(a)], we find that $\rho(\ell,t)$ approximately collapses for 
$z = 1$. Moreover, for the longest time $t = 100$ shown here, 
$\rho(\ell,t)$ exhibits a flat plateau around $\ell = L/2$ at the saturation 
value $\rho(\ell,t) = 0.75$, which indicates full scrambling within this area. 
In the latter case [Fig.\ \ref{Fig_Rho}~(b)], we find that $\rho(\ell,t)$ is 
similar to the density profiles $\langle Z_\ell(t)\rangle$ in Fig.\ 
\ref{Fig_Hydro}~(c), with a Gaussian shape in the bulk and additional heavy 
tails. Moreover, as in Fig.\ \ref{Fig_Hydro}~(c), we find a data
collapse for $z \approx 1.85$. As discussed in the main text, 
this similarity of $\langle Z_\ell(t)\rangle$ and $\rho(\ell,t)$ is expected in 
U$(1)$-symmetric circuits since ${\cal O}(t)$ will quickly be 
dominated by $Z$ operators such that the operator spreading will be impacted by 
the hydrodynamic behavior of the conserved quantity. Moreover, we note that the 
value of $z \approx 1.85$ seems consistent with the 
growth of entanglement discussed in Fig.\ 4, where we
found deviations from the theoretically expected $z = 2$ at $\alpha = 3$. We 
here leave it to future work to study finite-size and finite-time 
effects in more details (but see 
\cref{Fig::finite_size,Fig::finite_size_full_Clifford}) and to analyze 
the potential impacts of the non-Gaussian tails on $S(t)$.

\begin{figure}[tb]
 \centering
 \includegraphics[width = 0.9\textwidth]{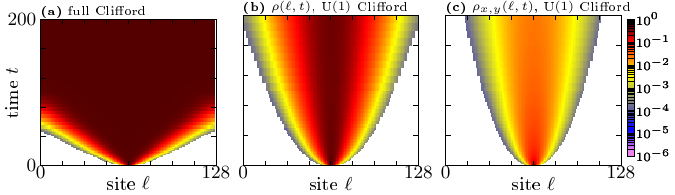}
 \caption{Operator spreading resulting from ${\cal O}(0) = X_{L/2}$ in Clifford 
circuits with $L = 128$ and $\alpha = 5$. {\bf (a)} $\rho(\ell,t)$ under  
full Clifford evolution without conservation law. {\bf (b)} $\rho(\ell,t)$ in 
U$(1)$-symmetric Clifford circuits. Note that data in panels (a) and (b) is 
analogous to data for $L = 1024$ shown in Figs.\ 3~(a) and (d) in 
the main text. {\bf (c)} $\rho_{x,y}(\ell,t)$ [Eq.\ \eqref{Eq::RhoXY}] 
in U$(1)$-symmetric circuits.} 
 \label{Fig::LC_XY}
\end{figure}

While the operator string ${\cal O}(t)$ will quickly 
be dominated by $Z$ operators in the case of U$(1)$-symmetric Clifford gates,
it is interesting to study how $X$ and $Y$ operators spread within ${\cal 
O}(t)$. Even though their overall weight decays as $\propto t^{-1/z}$ [see 
inset in Fig.\ 3~(f) in the main text] their dynamics 
might differ from the hydrodynamic behavior of the $Z$ operators. To 
this end, 
we analyze in Fig.\ \ref{Fig::LC_XY}~(c) the spreading of $X$ and $Y$ operators 
in terms of the quantity, 
\begin{equation}\label{Eq::RhoXY}
 \rho_{x,y}(\ell,t) = \sum_{\sigma=x,y} \rho_\sigma(\ell,t) = 
 \sum_{\sigma=x,y} \text{tr}[{\cal 
O}_\ell(t)\Sigma^\sigma]/2\ ,  
\end{equation}
where, in contrast to $\rho(\ell,t)$, the sum now runs only over $X$ and $Y$ 
operators. [As already stated in the main text, ${\cal O}_\ell(t)$ denotes the 
local Pauli or identity matrix at the $\ell$th position of the operator string 
and $\Sigma^{\sigma = x,y,z} = \lbrace X,Y,Z\rbrace$. Given the orthogonality 
of Pauli matrices, we have $\text{tr}({\cal O}_\ell(t)\Sigma^\sigma)/2 = 
\delta_{{\cal 
O}_\ell(t),\Sigma^\sigma}$.] Focusing on $\alpha = 
5$, we find in Fig.\ \ref{Fig::LC_XY}~(c) that $\rho_{x,y}(\ell,t)$ spreads 
diffusively and is qualitatively very similar to our results for $\rho(\ell,t)$ 
in U$(1)$-symmetric circuits in the main text. This is highlighted by comparing 
to Fig.\ 
\ref{Fig::LC_XY}~(b), where we show $\rho(\ell,t)$ (i.e., including $Z$ 
operators) for the same system size $L = 128$. Thus, it appears that not only 
is the overall number of $X$ and $Y$ operators reduced [cf.\ Fig.\ 
3~(f)], but the spreading of $X$ and $Y$ operators is also 
affected by the presence of the conservation law. This fact is further 
emphasized by comparing the results for $\rho_{x,y}(\ell,t)$ in Fig.\ 
\ref{Fig::LC_XY}~(c) to the case of full Clifford circuits without conservation 
law in Fig.\ \ref{Fig::LC_XY}~(a) (here $\rho(\ell,t)$ and $\rho_{x,y}(\ell,t)$ 
are equivalent as no operator is favored), where the spreading is ballistic.
Let us note that this slow spreading of $X$ and $Y$ operators in 
U$(1)$-symmetric Clifford circuits shown in Fig.\ \ref{Fig::LC_XY}~(c) appears 
to differ to more generic Haar-random circuits with conservation law, 
where conserved operators yield a slow hydrodynamic 
bulk that lags behind the significantly faster spreading front dominated by 
nonconserved operators \cite{Khemani2018S}. 

\begin{figure}[tb]
\centering
\includegraphics[width=0.8\textwidth]{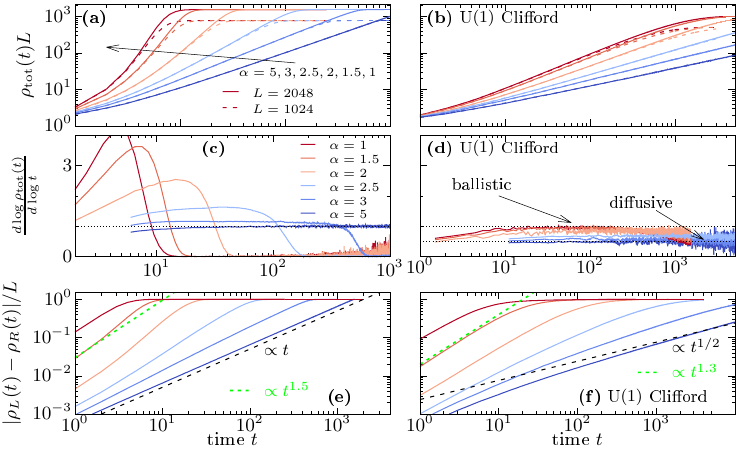}
\caption{{\bf (a)} Total support $\rho_\text{tot}(t)L$ of ${\cal O}(t)$ for $L 
= 1024,2048$ and different values of $\alpha$ (arrow), resulting from isolated 
operator ${\cal O}(0) = X_{L/2}$ under full Clifford evolution without 
conservation law. Note that data is analogous to Fig.\ 3~(c) in the 
main text. {\bf (c)} Logarithmic derivative $d \log \rho_\text{tot}(t)/d\log t$ 
of the data in panel (a). {\bf (e)} Normalized difference $|\rho_L(t) - 
\rho_R(t)|/L$ between left and right endpoint of ${\cal O}(t)$, cf.\ Eq.\ 
\eqref{Eq::LR}. Panels {\bf (b)}, {\bf (d)}, and {\bf (f)} show analogous data, 
but now for Clifford circuits with U$(1)$-symmetric gates.}  
\label{Fig_Density}
\end{figure}

Let us now provide some additional analysis regarding the dynamics of the total 
support $\rho_\text{tot}(t)$, already considered in Figs.\ 
3~(c) and (f) in the main text. In particular, in Figs.\ 
\ref{Fig_Density}~(a) and (b), we show $\rho_\text{tot}(t)$ for different 
$\alpha$ values in circuits without and with conservation law and two system 
sizes $L = 1024,2048$. Comparing the curves for different $L$, we find that 
especially for larger values of $\alpha$, finite-size effects are 
well-controlled on the time scales shown here. To study the growth of 
$\rho_\text{tot}(t)$ in more detail, Figs.\ \ref{Fig_Density}~(c) and (d) show 
its logarithmic derivative. While for large $\alpha = 5$ we recover ballistic 
behavior ($d \log \rho_\text{tot}(t)/d \log t \to 1$) in the unsymmetric case 
and diffusive growth ($d \log \rho_\text{tot}(t)/d \log t \to 0.5$) in 
U$(1)$-symmetric circuits, we find that the two gate sets behave 
drastically different at lower $\alpha$. Specifically, for unsymmetric circuits 
with $\alpha \lesssim 2$ we are unable to find an extended window with constant 
 $d \log \rho_\text{tot}(t)/d \log t$, indicating that $\rho_\text{tot}(t)$ is 
not described by a power law anymore. While this seems consistent with the 
phase diagram for long-range systems obtained in 
\cite{Zhou2020S}, where it was argued that the nature of operator spreading 
changes for $\alpha \leq 2$, we cannot exclude the impact of 
finite-size/finite-time effects which are clearly more pronounced for smaller 
$\alpha$. 
Interestingly, in contrast to full Clifford evolution, we find that the growth 
of $\rho_\text{tot}(t)$ in U$(1)$-symmetric circuits appears to be described by 
a power law for all values of $\alpha$ shown here, with approximately 
constant $d \log \rho_\text{tot}(t)/d \log t$ over an extended time window 
[Fig.\ \ref{Fig_Density}~(d)]. Surprisingly, however, we find that the growth 
of $\rho_\text{tot}(t)$ never exceeds ballistic $\propto t$ behavior even for 
small $\alpha$, which is somewhat unexpected as the dynamical transport 
exponent $z$ continues to decrease 
for $\alpha < 2$, cf.\ Fig.\ 1~(b) in the main text.  

In addition to the total support $\rho_\text{tot}(t)$ of ${\cal O}(t)$, it is 
instructive to study the left and right endpoints $\rho_L(t)$, $\rho_R(t)$ of 
the operator string, see also Fig.~4 in the main text, 
\begin{equation}\label{Eq::LR}
 \rho_L(t) = \min \lbrace \ell\ |\ \text{tr}[{\cal O}_\ell(t) 
\Sigma^{x,y,z}] 
\neq 0 \rbrace\ ,\quad  \rho_R(t) = \max \lbrace \ell\ |\ \text{tr}[{\cal 
O}_\ell(t) 
\Sigma^{x,y,z}] 
\neq 0 \rbrace\ , 
\end{equation}
which are defined as the smallest and largest lattice site $\ell$, for which  
${\cal O}(t)$ is a non-identity Pauli matrix. For the initial condition ${\cal 
O}(0)= X_{L/2}$, we obviously have $\rho_L(t) = \rho_R(t) = L/2$. For $t > 0$, 
the difference $|\rho_L(t) - \rho_R(t)|$ is expected to grow. In Figs.\ 
\ref{Fig_Density}~(e) and (f), $|\rho_L(t) - \rho_R(t)|/L$ is shown for 
circuits without and with conservation law and different values of $\alpha$. 
While for $\alpha = 5$ we recover ballistic $\propto t$ or diffusive $\propto 
t^{1/2}$ behavior respectively, the growth of $|\rho_L(t) - \rho_R(t)|$ appears 
to become more and more similar for smaller $\alpha$. Specifically, considering 
$\alpha = 1.5$, we approximately find $|\rho_L(t) - \rho_R(t)| \propto t^{1.5}$ 
for unsymmetric circuits, while $|\rho_L(t) - \rho_R(t)| \propto t^{1.3}$ for 
circuits with U$(1)$ symmetry. While we should note that it is rather 
tricky to fit a power law given the short time scales for such small $\alpha$, 
the overall behavior in Figs.\ \ref{Fig_Density}~(e) and (f) appears at least 
consistent with our observation that entanglement dynamics $S(t)$ becomes 
almost unaffected by the presence of a conservation law once $\alpha$ is 
sufficiently small, 
cf.\ Fig.\ 4 in the main text.   

Comparing the data of $\rho_\text{tot}(t)$ in Figs.\ 
\ref{Fig_Density}~(a) and (b) with the results for $|\rho_L(t) - \rho_R(t)|$ in 
Figs.\ \ref{Fig_Density}~(e) and (f), we conclude that although the shape of 
the light cones may become similar for circuits with and without conservation 
law if $\alpha$ is small, the interior of the light cone behaves notably 
different. In particular, the comparatively slower growth of 
$\rho_\text{tot}(t)$ in U$(1)$-symmetric circuits suggests that the 
operator string ${\cal O}(t)$ still contains a larger fraction of identity 
operators [and as we discussed in Fig.\ 3~(f), many more $Z$ 
operators than $X$ and $Y$], whereas in circuits without conservation law one 
quickly approaches the equilibrium distribution where $X$, $Y$, $Z$, and 
identity operators all occur with probability $1/4$. 

\subsection{Finite-size scaling of the dynamical exponent $z$}
\label{sec:finite_size_scaling_appendix}

In this section we explain the procedure we use for estimating the magnitude of
finite-size effects in our estimates of the dynamical critical exponent $z$
from the entanglement data. This is necessitated by the fact that we observe
fairly significant drifts with system size in our estimates of $z$, even at the
large systems sizes $L \sim 10^{3}$ accessible with Clifford circuits. In
passing, we note that it may be the case that Clifford circuits exhibit larger
finite-size effects than less structured models, such as Haar-random circuits.
It would be interesting to understand better whether the much larger system
sizes accessible with Clifford circuits compensate for any larger propensity
for significant finite-size effects. 

To extract $z$ from the entanglement growth, first for each system size $L$ we
calculate the `saturation time' $t_{\mathrm{sat}}$, which for practical
purposes we define to be the smallest time at which $S_{L/2}(t)$ is within 1\%
of its steady-state value. Assuming the entanglement growth is dominated by the
asymptotic scaling $S(t) \sim t^{1/z}$---which empirically is what we
observe---and using the fact that the steady-state entanglement is
$\mathcal{O}(L)$, the saturation time should scale as $t_{\mathrm{sat}} \sim
L^{z}$. Thus for a given set of system sizes $\{L_{i}\}$, we can estimate $z$
using a linear fit of $\log{t_{\mathrm{sat}}}$ vs $\log{L}$.

To account for finite-size effects, we artificially restrict the dataset we use
for this fit to system sizes $L \leq L_{\mathrm{end}}$ for some maximum
$L_{\mathrm{end}}$, obtaining a corresponding estimate $z(L_{\mathrm{end}})$.
We then vary $L_{\mathrm{end}}$ from $L_{\mathrm{end}} = 128$ to the largest
system size available, typically $L_{\mathrm{end}} = 1024$ or $2048$ depending
on the value of $\alpha$. 

Having obtained a range of estimates $\{z(L_{\mathrm{end}})\}$ for different
values of $L_{\mathrm{end}}$, we perform an initial fit of these estimates to a
power-law, $z(L_{\mathrm{end}}) = b(x_{0} + 1/L_{\mathrm{end}})^{-a}$. In
principle this initial fit can be used to perform the extrapolation of $z$ as
$1/L_{\mathrm{end}} \to 0$. However, this may give undue weight to this
particular power-law fit, since the extrapolated value at 0 can depend somewhat
sensitively on the parameters of the fit. To account for this, we perform a
form of `least squares Monte Carlo', over a parameter space centered around the
parameters obtained from the initial fit. We randomly draw a set of parameters
$\mathbf{p}=(a,b,x_{0})$ with probability proportional to the 
inverse square of the
least squares cost function $\epsilon(\mathbf{p}) =
\sum_{i}(y_{\mathbf{p}}(x_{i}) - y_{i})^{2} / \sigma_{i}^{2}$, where
$y_{\mathbf{p}}(x_{i}) = b(x_{0} + x_{i})^{-a}$, and the data $(x_{i}, y_{i},
\sigma_{i})$ are the values of $1/L_{\mathrm{end}}$, the corresponding
estimates $z(L_{\mathrm{end}})$, and their errors. For each set of parameters
$\mathbf{p}$ we obtain an estimate of the extrapolated value at
$1/L_{\mathrm{end}} = 0$. To get our final estimate of $z$, we draw $10^{5}$
Monte Carlo samples, and take the median extrapolated value. The lower and
upper error bars are given by the values of $z$ at which $2.5\%$ and $97.5\%$
respectively of the extrapolated values are below these thresholds.

\begin{figure}[t]
	\centering
	\includegraphics[width=0.9\columnwidth]{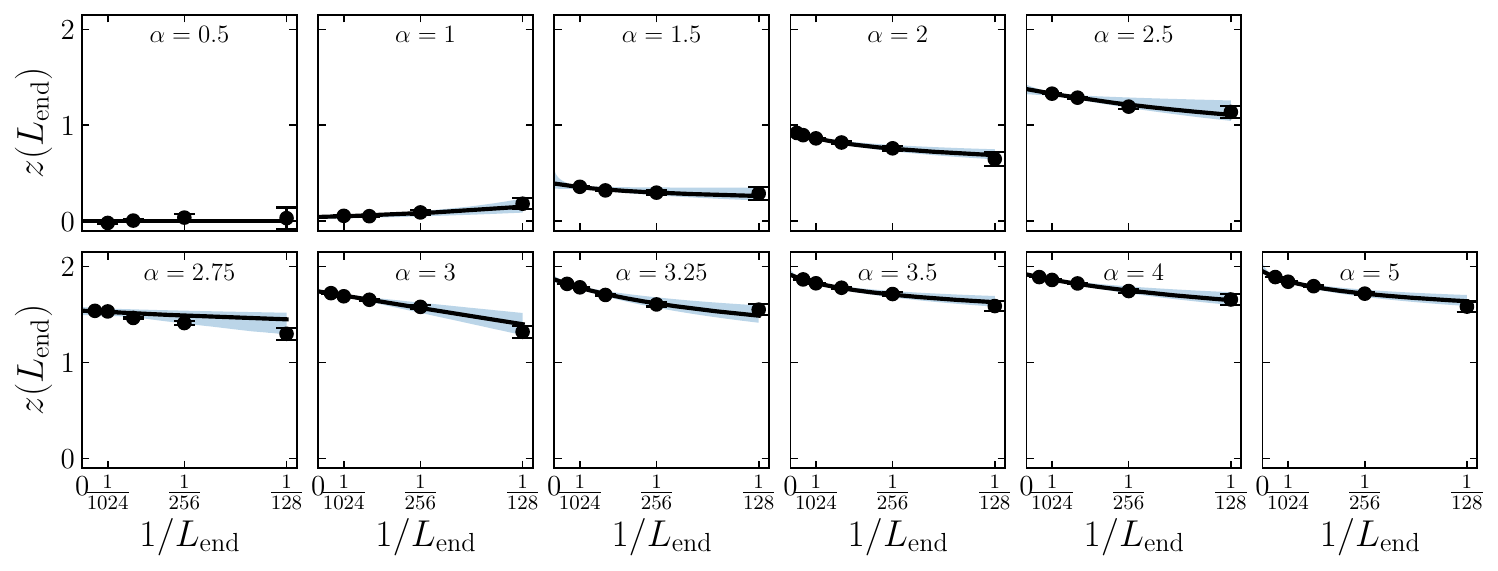}
	\caption{Extrapolation procedure for the dynamical critical exponent $z$
		in U$(1)$-symmetric Clifford circuits, for different values of 
$\alpha$.
		Data points come from estimates using a linear fit of
		$\log{t_{\mathrm{sat}}}$ vs $\log{L}$, while the black line is 
the median
		fit from the `least squares Monte Carlo' procedure described in 
the main
		text, and the shaded blue area is the corresponding $95\%$ 
confidence
		interval. In most cases the trend as $1/L_{\mathrm{end}} \to 0$ 
is for the
		$z$ estimate to increase, except at $\alpha = 1$, where the 
estimate
		decreases towards zero, and $\alpha = 0.5$, where the estimate 
remains very
		close to zero.}
	\label{Fig::finite_size}
\end{figure}

\begin{figure}[tb]
	\centering	
\includegraphics[width=0.7\columnwidth]{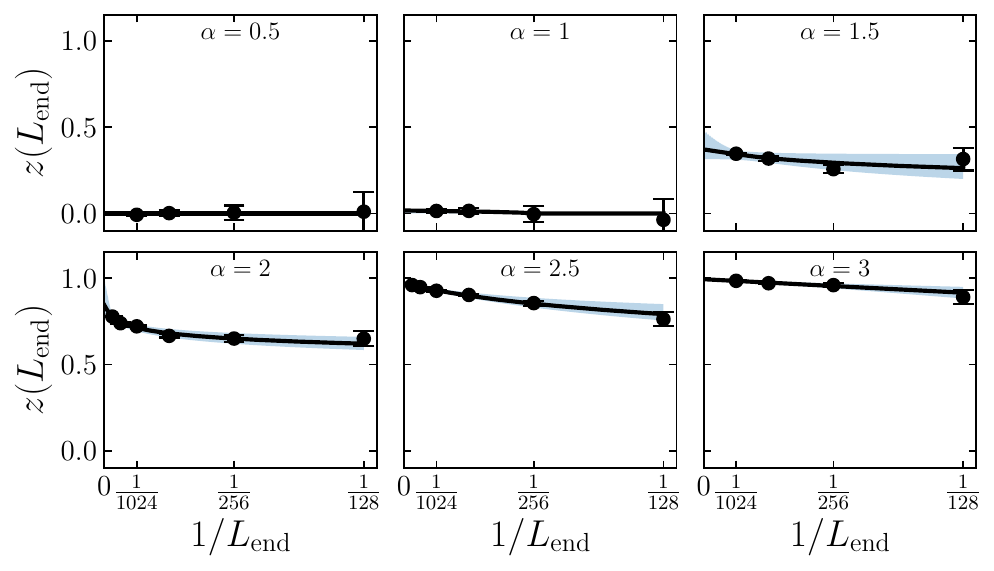}
	\caption{Extrapolation procedure for the dynamical critical exponent $z$
		in asymmetric Clifford circuits, for different values of 
$\alpha$.
		Data points come from estimates using a linear fit of
		$\log{t_{\mathrm{sat}}}$ vs $\log{L}$, while the black line is 
the median
		fit from the `least squares Monte Carlo' procedure described in 
the main
		text, and the shaded blue area is the corresponding $95\%$ 
confidence
		interval.}
	\label{Fig::finite_size_full_Clifford}
\end{figure}

These extrapolation procedures are shown in 
\cref{Fig::finite_size,Fig::finite_size_full_Clifford} for U$(1)$-symmetric and 
asymmetric Clifford circuits respectively. In most cases the trend as 
$1/L_{\mathrm{end}} \to 0$ is for the $z$ estimate to increase, except for 
U$(1)$-symmetric circuits at $\alpha = 1$, where the estimate decreases towards 
zero, and for the same circuits at $\alpha = 0.5$, where the estimate remains 
very close to zero. We note that the uncertainty in the estimate 
seems to be larger around $\alpha \approx 3$, where the transition occurs from 
short- to long-range behavior. As discussed in the main text, this may be 
related to the fact that both $\langle Z_\ell(t) \rangle$ 
and $\rho(\ell,t)$ develop heavy non-Gaussian tails at $\alpha \approx 3$ 
(see \cref{Fig_Hydro}). We defer further investigation of this relationship to 
future work.
\subsection{Comparison between long-ranged circuits and circuits with 
nearest-neighbor gates}

\begin{figure}[b]
\centering
\includegraphics[width=0.7\textwidth]{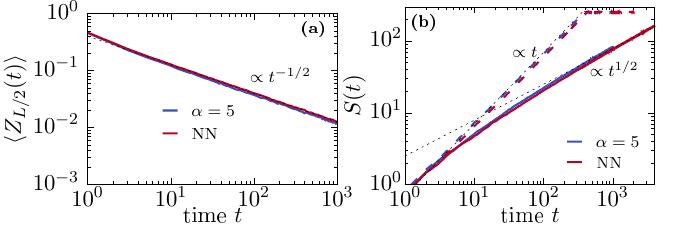}
\caption{{\bf (a)} $\langle Z_{L/2}(t) \rangle$ for long-range circuits with 
$\alpha = 5$ and circuits with nearest-neighbor (NN) gates, obtained 
analogous to Fig.\ 2 in the main text. {\bf (b)} 
Corresponding entanglement growth in circuits with U$(1)$ conservation law 
(solid curves), resulting from the initial state $\ket{\rightarrow}^{\otimes 
L}$. As a comparison, we also show $S(t)$ for circuits without 
conservation law (dashed curves). In all cases, we have $L = 512$ 
and periodic boundary conditions.}
\label{Fig::NN}
\end{figure}

For sufficiently large $\alpha$, the properties of circuits with long-range 
interactions approach those of strictly local circuits. This fact 
is demonstrated in Fig.\ \ref{Fig::NN}, where we compare transport and 
entanglement growth in long-range circuits with $\alpha = 5$ and circuits 
with 
nearest-neighbor gates. 
As shown in Fig.\ \ref{Fig::NN}~(a), the spin excitation $\langle 
Z_{L/2}(t)\rangle$ decays diffusively $\propto t^{-1/2}$ for both circuit 
variants and the data for the two different circuits agree very well with 
each other. Likewise, the entanglement production in long-range circuits with 
$\alpha = 5$ is essentially equivalent to that in local circuits. In 
particular, we find that $S(t)$ growth diffusively $\propto t^{1/2}$  
at long times for circuits with U$(1)$ conservation law, and linearly $\propto 
t$ for circuits without charge conservation.  

\subsection{Comparison between random Clifford circuits and Haar-random 
circuits}
\begin{figure}[b]
\centering
\includegraphics[width = 0.75\textwidth]{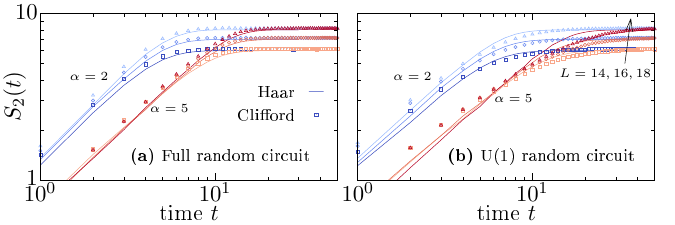}
\caption{Entanglement $S(t)$ for $\alpha = 2$ and 
$\alpha = 5$ in circuits {\bf 
(a)} without conservation law and {\bf (b)} with U$(1)$-symmetric gates. 
We 
compare the case of random Clifford circuits (symbols)
to circuits where 
the two-site gates are randomly drawn according to the Haar measure (curves). 
Data is shown for circuits with $L = 14,16,18$ and periodic boundaries. While 
for Clifford circuits all R\'enyi entropies are equivalent, we show $S_2(t)$ 
for Haar-random circuits.}
\label{Fig_Haar}
\end{figure}

Let us briefly compare entanglement dynamics in Clifford circuits 
to the case of Haar-random circuits. Such a 
comparison is shown in Fig.\ \ref{Fig_Haar}~(a) for circuits without 
conservation law and in Fig.\ \ref{Fig_Haar}~(b) for circuits with 
U$(1)$-symmetric gates. We here focus on $\alpha = 2$ and $\alpha = 5$ and show 
data for three different systems sizes $L = 14,16,18$ (note that in contrast to 
Clifford circuits, the simulation of Haar-random circuits is exponentially 
costly in $L$). While for Clifford circuits all R\'enyi entropies are 
equivalent, we show $S_2(t)$ in the case of Haar-random gates. 

On one hand, for the case without conservation law, we find that 
the entanglement dynamics is very similar for Clifford and Haar-random 
circuits, which emphasizes the fact that random Clifford circuits form 
unitary $2$-designs and can imitate the properties of more 
generic types of unitary evolution. On the other hand, 
in the U$(1)$-symmetric case, we find that the dynamics of $S(t)$ is again 
qualitatively similar in Clifford and Haar-random circuits, but $S(t)$ appears 
to saturate slightly faster towards its steady-state value for Haar-random 
gates. As already discussed in the context of Eqs.\ 
\eqref{Eq::Cliff12}-\eqref{Eq::Cliff12_2}, we attribute this difference to the 
fact that the set of U$(1)$-symmetric Clifford gates contains a 
comparatively high percentage of separable gates that generate no entanglement. 
We expect, however, that this will not change the dynamical critical exponent 
$z$, but only affect the coefficient of the $S(t) \propto t^{1/z}$ scaling. In 
particular, in the thermodynamic limit $L \to \infty$, we expect that the 
dynamics of $S(t)$ in U$(1)$-symmetric 
Clifford circuits at finite times is representative of the dynamics of higher 
R\'enyi entropies $S_{n>1}(t)$ in more generic quantum many-body systems.

\subsection{Entanglement dynamics in long-range quantum many-body systems}

\begin{figure}[tb]
\centering
\includegraphics[width=0.45\textwidth]{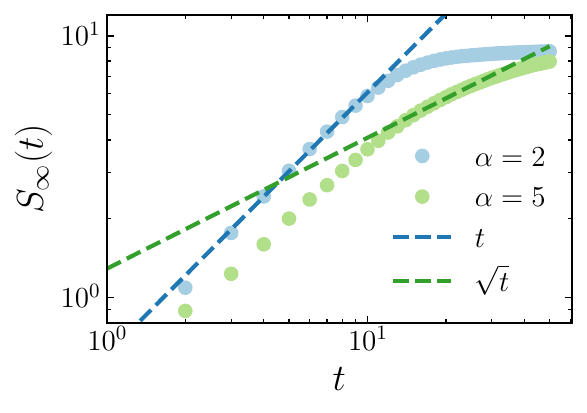}
\caption{$S_{\infty}(t)$ for different $\alpha$ in the 
long-range tilted field Ising model with $L=22$, showing similar behavior to 
the 
Clifford circuits. The results are averaged over initial product states, but 
sample-to-sample fluctuations are small.}
\label{Fig::Ising}
\end{figure}

To further support the generality of our conclusions of Fig.~4 in the main 
text, let us now consider an actual chaotic quantum many-body system. 
Specifically, we show in Fig.\ \ref{Fig::Ising} the 
growth of the R\'{e}nyi-$\infty$ entropy 
$S_{\infty}(t)$ of the long-range tilted field Ising 
model~\cite{Kim2013S, Rakovszky2019S}, with 
\begin{equation}\label{Eq::Ising}
 H = \sum_{i < j} \frac{J}{|i-j|^{\alpha^{\prime}}} Z_{i} Z_{j} + \sum_{i} 
h_{z} Z_{i} + h_{x} X_{i} - 
J(Z_{1} + Z_{L})\ , 
\end{equation} 
where $\alpha^{\prime} = \alpha/2$ to allow comparison 
with 
the random circuits~\cite{Zhou2020S, Block2022S}, 
and we set $J=1, h_{z} = 0.8090, 
h_{x} = 0.9045$. Although the results are limited by small system sizes 
($L=22$), the overall behavior is similar to the Clifford circuits: for 
$\alpha=5$ there is a crossover from $t$ to $\sqrt{t}$ growth at long times, 
while for $\alpha = 2$ the growth remains $\propto t$ before saturating, 
consistent with Eq.\ (2) in the main text. This suggests that our findings for 
Clifford circuits can carry over to chaotic many-body quantum systems with 
long-range interactions. Furthermore, the data in Fig.\ \ref{Fig::Ising} 
demonstrates that constrained entanglement growth occurs not only for models 
with charge conservation (such as in the main text), but applies also more 
generally for other conservation laws \cite{Rakovszky2019S}. In fact, the Ising 
model in Eq.\ \eqref{Eq::Ising} does not conserve the total magnetization, but 
only has total energy as a conserved quantity and exhibits diffusive energy 
transport in the short-range limit \cite{Kim2013S}.    

\subsection{Two-dimensional circuits}

Given the efficient simulability of Clifford circuits, 
we are able to study transport and entanglement growth also in two-dimensional 
long-range quantum systems, which would otherwise be challenging even for 
state-of-the-art numerical techniques. We consider circuits with 
square geometry and total qubit number $L^2 = L \times L$. Similar to  
the 1$d$ case discussed in the main text, a single time step is defined as the 
application of $L^2$ random two-qubit gates of range $r$, drawn according to 
the 
probability distribution $P(r) \propto r^{-\alpha}$. Specifically, for two 
qubits at 
positions $(x_1,y_1)$ and $(x_2,y_2)$, we define $r$ as 
\begin{equation}
 r = |x_1 - x_2| + |y_1 - y_2|\ . 
\end{equation}

\begin{figure}[b]
\centering
\includegraphics[width=0.9\textwidth]{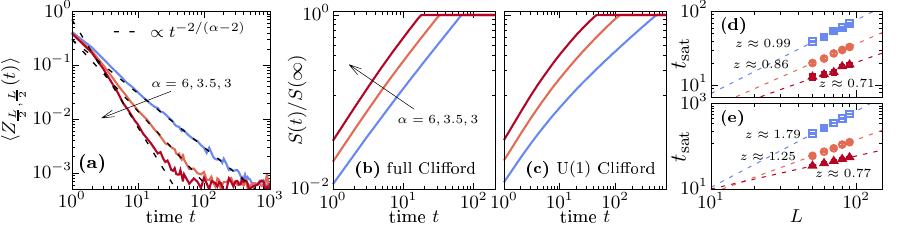}
\caption{Transport and 
entanglement growth in 2$d$ lattices. {\bf 
(a)} $\langle Z_{\frac{L}{2},\frac{L}{2}}(t)\rangle$ for different 
$\alpha$ and $L = 40$. Dashed lines indicate power-laws $\propto 
t^{-d/(\alpha-d)}$. {\bf (b)} $S(t)$ for a half system bipartition under full 
Clifford evolution with $\alpha = 
6,3.5,3$ in $90\times 90$ lattice with periodic boundaries. {\bf (c)} 
Analogous to (b), but 
now for Clifford
circuits with U$(1)$ conservation law. [{\bf (d),(e)}] Saturation time 
$t_\text{sat}$ of $S(t)$ versus $L\in[50,90]$ for full and U$(1)$-symmetric 
circuits. The exponent $z$ of the fitted power-law scaling $t_\text{sat}\propto 
L^z$ is indicated next to the data. For $\alpha \lesssim 3$, the scaling in 
circuits 
with and without conservation law is found to become similar.}
\label{Fig::2D}
\end{figure}

Analogous to our analysis in Fig.\ 2, we study transport by 
considering the U$(1)$-symmetric Clifford evolution of an isolated $Z$ operator 
initially defined at the central site of the 2$d$ lattice, ${\cal 
U}Z_{\frac{L}{2},\frac{L}{2}}{\cal 
U}^\dagger$. Crucially, the picture of long-range random walks of the single 
$Z$ 
operator 
discussed in the main text 
generalizes 
directly to higher-dimensional lattices, i.e., for a single circuit realization 
the operator will always remain of length one. For the circuit-averaged dynamics
of the $Z$ excitation, the U$(1)$ conservation law then leads to a 
power-law decay, $\langle Z_{\frac{L}{2},\frac{L}{2}}(t)\rangle \propto 
t^{-d/z}$, which reflects 
the 
different $\alpha$-dependent hydrodynamic regimes. In particular, based on the 
prediction from L\'evy flights (see \cite{Schuckert2020S}), we expect diffusion 
($z = 2$) 
for $\alpha \geq d + 2$, while $z = \alpha - d$ for $d < \alpha \leq d + 
2$, and a  
breakdown of hydrodynamics for $\alpha < d$. For $d = 2$, this yields 
\begin{equation}\label{Eq::z_2d}
 z = \begin{cases}
                                      2, \ &\alpha \geq 4 \\
                                      \alpha-2, \ &2 < \alpha \leq 4  
                                    \end{cases}\ . 
\end{equation}
Let us emphasize again that for Hamiltonian systems with coupling constant $J 
\propto r^{-\alpha'}$, these bounds have to be rescaled according to $\alpha = 
2\alpha'$ \cite{Zhou2020S, Schuckert2020S}.
Focusing on $\alpha = 6,3.5,3$, Fig.\ \ref{Fig::2D}~(a) unveils a 
convincing agreement of our numerics with Eq.\ \eqref{Eq::z_2d}, where we 
consider $\langle Z_{\frac{L}{2},\frac{L}{2}}(t)\rangle$ in circuits of 
size $40\times 40$. 
Note that for smaller $\alpha$, the analysis becomes more difficult due to 
finite-size effects. 

We now turn to entanglement dynamics in 2$d$ circuits. To this end, 
we consider circuits with periodic boundary conditions and calculate $S(t)$ for 
a half-system bipartition. In Figs.\ \ref{Fig::2D}~(b) and (c), we show $S(t)$ 
for circuits without conservation 
law as well as for U$(1)$-symmetric Clifford circuits, where we 
again focus on $\alpha = 6,3.5,3$. As expected, the growth of 
$S(t)$ is notably slower in U$(1)$-symmetric circuits due to the constraint 
imposed by the conservation law. However, as was already discussed in Ref.\ 
\cite{Znidaric2020S}, we find that it is actually rather difficult to observe 
the conjectured asymptotic scaling $S(t) \propto t^{1/z}$. Therefore, we here 
proceed analogous to our analysis in the context of Fig.\ 4 
and extract the saturation time $t_\text{sat}$ of $S(t)$ for circuit sizes 
ranging from $50\times 50$ to $90\times 90$, see Figs.\ \ref{Fig::2D}~(d) and 
(e). [See labels next to the data in Figs.\ \ref{Fig::2D}~(d) and (e) for the 
obtained values of $z$.] In particular, for circuits without conservation law 
and $\alpha = 6$, we recover the expected linear growth of $S(t)$ with $z 
\approx 1$. In contrast, for U$(1)$-symmetric 
circuits, we find a substantially larger value $z \approx 1.79$, which is 
however 
smaller than the conjectured value $z = 2$. The deviation from the diffusive 
value may be 
due to finite-size effects \cite{Znidaric2020S}, and we note that similar 
deviations also occurred in 1$d$ circuits discussed in Fig.\ 
4. Interestingly, for $\alpha = 3$, we find that the 
scaling of $t_\text{sat}$ becomes again rather similar for both circuit 
variants. Analogous to our discussion of 1$d$ circuits in the main text, 
this finding can be understood due to the fact that transport becomes 
sufficiently fast for $\alpha \leq 3$ in 2$d$, such that the presence
of the hydrodynamic mode becomes less and less relevant.


\begin{thebibliography}{99}

\bibitem{dalessio2016}
L. D'Alessio, Y. Kafri, A. Polkovnikov, and M. Rigol,
{\it From Quantum Chaos and Eigenstate Thermalization to Statistical Mechanics 
and Thermodynamics},
Adv. Phys. {\bf 65}, 239 (2016).

\bibitem{nandkishore2015}
R. Nandkishore and D. A. Huse,
{\it Many-Body Localization and Thermalization in Quantum Statistical 
Mechanics},
Annu. Rev. Condens. Matter Phys. {\bf 6}, 15 (2015).

\bibitem{Bertini2021}
B. Bertini, F. Heidrich-Meisner, C. Karrasch, T. Prosen, R. Steinigeweg, and M. 
\v{Z}nidari\v{c},
{\it Finite-temperature transport in one-dimensional quantum lattice models},
Rev. Mod. Phys. {\bf 93}, 025003 (2021). 

\bibitem{Li2017}
J. Li, R. Fan, H. Wang, B. Ye, B. Zeng, H. Zhai, X. Peng, and J. Du, 
{\it Measuring Out-of-Time-Order Correlators on a Nuclear Magnetic Resonance 
Quantum Simulator},
Phys. Rev. X {\bf 7}, 031011 (2017). 

\bibitem{Gaerttner2017}
M. G\"arttner, J. G. Bohnet, A. Safavi-Naini, M. L. Wall, J. J. Bollinger, and 
A. M. Rey, 
{\it Measuring out-of-time-order correlations and multiple quantum spectra in a 
trapped ion quantum magnet},
Nat. Phys. {\bf 13}, 781 (2017). 

\bibitem{Landsman2019}
K. A. Landsman, C. Figgatt, T. Schuster, N. M. Linke, B. Yoshida, N. Y. Yao, 
and C. Monroe,
{\it Verified quantum information scrambling},
Nature {\bf 567}, 61 (2019). 

\bibitem{Blok2021}
M. S. Blok, V. V. Ramasesh, T. Schuster, K. O'Brien, J. M. Kreikebaum, D. 
Dahlen, A. Morvan, B. Yoshida, N. Y. Yao, and I. Siddiqi,
{\it Quantum Information Scrambling on a Superconducting Qutrit Processor},
Phys. Rev. X {\bf 11}, 021010 (2021). 

\bibitem{Haegeman2011}
J. Haegeman, J. I. Cirac, T. J. Osborne, I. Pi\v{z}orn, H. Verschelde, and F. 
Verstraete,
{\it Time-Dependent Variational Principle for Quantum Lattices},
Phys. Rev. Lett. {\bf 107}, 070601 (2011). 

\bibitem{Paeckel2019}
S. Paeckel, T. K\"ohler, A. Swoboda, S. R. Manmana, U. Schollw\"ock, and C. 
Hubig, 
{\it Time-evolution methods for matrix-product states}, 
Ann. Phys. {\bf 411}, 167998 (2019). 

\bibitem{Rakovszky2022}
T. Rakovszky, C. W. von Keyserlingk, and F. Pollmann,
{\it Dissipation-assisted operator evolution method for capturing hydrodynamic 
transport},
Phys. Rev. B {\bf 105}, 075131 (2022).

\bibitem{White2018}
C. D. White, M. Zaletel, R. S. K. Mong, and G. Refael, 
{\it Quantum dynamics of thermalizing systems}, 
Phys. Rev. B {\bf 97}, 035127 (2018). 

\bibitem{Wurtz2018}
J. Wurtz, A. Polkovnikov, and D. Sels, 
{\it Cluster truncated Wigner approximation in strongly interacting systems}, 
Ann. Phys. {\bf 395}, 341 (2018).

\bibitem{Heitmann2020}
T. Heitmann, J. Richter, D. Schubert, and R. Steinigeweg, 
{\it Selected applications of typicality to real-time dynamics of quantum 
many-body systems}, 
Z. Naturforsch. A {\bf 75}, 421 (2020). 

\bibitem{Nahum2017}
A. Nahum, J. Ruhman, S. Vijay, and J. Haah,
{\it Quantum Entanglement Growth under Random Unitary Dynamics}, 
Phys. Rev. X {\bf 7}, 031016 (2017).

\bibitem{Keyserlingk2018}
C. W. von Keyserlingk, T. Rakovszky, F. Pollmann, and S. L. Sondhi,
{\it Operator Hydrodynamics, OTOCs, and Entanglement Growth in Systems without 
Conservation Laws}, 
Phys. Rev. X {\bf 8}, 021013 (2018).

\bibitem{Nahum2018}
A. Nahum, S. Vijay, and J. Haah, 
{\it Operator Spreading in Random Unitary Circuits}, 
Phys. Rev. X {\bf 8}, 021014 (2018).

\bibitem{Brandao2021}
F. G. S. L. Brand$\tilde{\text{a}}$o, W. Chemissany, N. Hunter-Jones, R. Kueng, 
and J. Preskill, 
{\it Models of Quantum Complexity Growth}, 
PRX Quantum {\bf 2}, 030316 (2021). 

\bibitem{Khemani2018}
V. Khemani, A. Vishwanath, and D. A. Huse, 
{\it Operator Spreading and the Emergence of Dissipative Hydrodynamics under 
Unitary Evolution with Conservation Laws}, 
Phys. Rev. X {\bf 8}, 031057 (2018). 

\bibitem{Rakovszky2018}
T. Rakovszky, F. Pollmann, and C. W. von Keyserlingk, 
{\it Diffusive Hydrodynamics of Out-of-Time-Ordered Correlators with Charge 
Conservation}, 
Phys. Rev. X {\bf 8}, 031058 (2018). 

\bibitem{Moudgalya2021}
S. Moudgalya, A. Prem, D. A. Huse, and A. Chan, 
{\it Spectral statistics in constrained many-body quantum chaotic systems}, 
Phys. Rev. Research {\bf 3}, 023176 (2021). 

\bibitem{Bertini2019}
B. Bertini, P. Kos, and T. Prosen,
{\it Exact Correlation Functions for Dual-Unitary Lattice Models in $1+1$ 
Dimensions}, 
Phys. Rev. Lett. {\bf 123}, 210601 (2019).

\bibitem{Claeys2021}
P. W. Claeys and A. Lamacraft, 
{\it Ergodic and Nonergodic Dual-Unitary Quantum Circuits with Arbitrary Local 
Hilbert Space Dimension}, 
Phys. Rev. Lett. {\bf 126}, 100603 (2021). 

\bibitem{Bertini2018}
B. Bertini, P. Kos, and T. Prosen, 
{\it Exact Spectral Form Factor in a Minimal Model of Many-Body Quantum Chaos}, 
Phys. Rev. Lett. {\bf 121}, 264101 (2018). 

\bibitem{Chan2018}
A. Chan, A. De Luca, and J. T. Chalker, 
{\it Spectral Statistics in Spatially Extended Chaotic Quantum Many-Body 
Systems}, 
Phys. Rev. Lett. {\bf 121}, 060601 (2018). 

\bibitem{Skinner2019}
B. Skinner, J. Ruhmann, and A. Nahum, 
{\it Measurement-Induced Phase Transitions in the Dynamics of Entanglement}, 
Phys. Rev. X {\bf 9}, 031009 (2019). 

\bibitem{Potter2021}
A. C. Potter and R. Vasseur, 
{\it Entanglement dynamics in hybrid quantum circuits}, 
arXiv:2111.08018.

\bibitem{Preskill2018}
J. Preskill, 
{\it Quantum Computing in the NISQ era and beyond}, 
Quantum {\bf 2}, 79 (2018). 

\bibitem{Richter2021}
J. Richter and A. Pal, 
{\it Simulating Hydrodynamics on Noisy Intermediate-Scale Quantum Devices with 
Random Circuits}, 
Phys. Rev. Lett. {\bf 126}, 230501 (2021). 

\bibitem{Lunt2021}
O. Lunt, J. Richter, and A. Pal, 
{\it Quantum simulation using noisy unitary circuits and measurements}, 
arXiv:2112.06682. 

\bibitem{Arute2019}
F. Arute {\it et al.}, 
{\it Quantum supremacy using a programmable superconducting processor}, 
Nature {\bf 574}, 505 (2019).

\bibitem{Mi2021}
X. Mi {\it et al.}, 
{\it Information scrambling in quantum circuits}, 
Science {\bf 374}, 1479 (2021). 

\bibitem{Lux2014}
J. Lux, J. M\"uller, A. Mitra, and A. Rosch, 
{\it Hydrodynamic long-time tails after a quantum quench}, 
Phys. Rev. A {\bf 89}, 053608 (2014). 

\bibitem{Bohrdt2017}
A. Bohrdt, C. B. Mendl, M. Endres, and M. Knap, 
{\it Scrambling and thermalization in a diffusive quantum many-body system}, 
New J. Phys. {\bf 19}, 063001 (2017). 

\bibitem{Richter2018}
J. Richter, F. Jin, H. De Raedt, K. Michielsen, 
J. Gemmer, and R. Steinigeweg, 
{\it Real-time dynamics of typical and untypical states in nonintegrable 
systems}, 
Phys. Rev. B {\bf 97}, 174430 (2018). 

\bibitem{Richter2019}
J. Richter, F. Jin, L. Knipschild, J. Herbrych, H. De Raedt, K. Michielsen, 
J. Gemmer, and R. Steinigeweg, 
{\it Magnetization and energy dynamics in spin ladders: Evidence of diffusion in 
time, frequency, position, and momentum}, 
Phys. Rev. B {\bf 99}, 144422 (2019). 

\bibitem{Kim2013}
H. Kim and D. A. Huse, 
{\it Ballistic Spreading of Entanglement in a Diffusive Nonintegrable System}, 
Phys. Rev. Lett. {\bf 111}, 127205 (2013). 

\bibitem{Rakovszky2019}
T. Rakovszky, F. Pollmann, and C. W. von Keyserlingk,
{\it Sub-ballistic Growth of R\'enyi Entropies due to Diffusion}, 
Phys. Rev. Lett. {\bf 122}, 250602 (2019). 

\bibitem{Huang2020}
Y. Huang, 
{\it Dynamics of R\'enyi entanglement entropy in diffusive qudit systems}, 
IOP SciNotes {\bf 1}, 035205 (2020).

\bibitem{Znidaric2020}
M. \v{Z}nidari\v{c},
{\it Entanglement growth in diffusive systems}, 
Commun. Phys. {\bf 3}, 100 (2020). 

\bibitem{Rakovszky2021}
T. Rakovszky, F. Pollmann, and C. von Keyserlingk, 
{\it Entanglement growth in diffusive systems with large spin}, 
Commun. Phys. {\bf 4}, 91 (2021).

\bibitem{Zhou2020_2}
T. Zhou and A. W. W. Ludwig, 
{\it Diffusive scaling of R\'enyi entanglement entropy}, 
Phys. Rev. Research {\bf 2}, 033020 (2020). 

\bibitem{Zhou2020}
T. Zhou, S. Xu, X. Chen, A. Guo, and B. Swingle,
{\it Operator L\'evy Flight: Light Cones in Chaotic Long-Range Interacting 
Systems}, 
Phys. Rev. Lett. {\bf 124}, 180601 (2020). 

\bibitem{Block2022}
M. Block, Y. Bao, S. Choi, E. Altman, and N. Y. Yao,
{\it Measurement-Induced Transition in Long-Range Interacting Quantum 
Circuits}, 
Phys. Rev. Lett. {\bf 128}, 010604 (2022). 

\bibitem{Harrow2009}
A. W. Harrow and R. A. Low, 
{\it Random Quantum Circuits are Approximate 2-designs}, 
Comm. Math. Phys. {\bf 291}, 257 (2009). 

\bibitem{Webb2015}
Z. Webb, 
{\it The Clifford group forms a unitary 3-design}, 
Quantum Inf. Comput. {\bf 16}, 1379 (2016).  

\bibitem{Saffman2010}
M. Saffman, T. G. Walker, K. M\o lmer, 
{\it Quantum information with Rydberg atoms}, 
Rev. Mod. Phys. {\bf 82}, 2313 (2010). 

\bibitem{Porras2004}
D. Porras and J. I. Cirac, 
{\it Effective Quantum Spin Systems with Trapped Ions}, 
Phys. Rev. Lett. {\bf 92}, 207901 (2004). 

\bibitem{Jurceviv2014}
P. Jurcevic, B. P. Lanyon, P. Hauke, C. Hempel, P. Zoller, R.
Blatt, and C. F. Roos, 
{\it Observation of entanglement propagation in a quantum many-body system}, 
Nature (London) {\bf 511}, 202 (2014).

\bibitem{Smith2016}
J. Smith, A. Lee, P. Richerme, B. Neyenhuis, P. W. Hess, P.
Hauke, M. Heyl, D. A. Huse, and C. Monroe, 
{\it Many-body localization in a quantum simulator with programmable random 
disorder}, 
Nat. Phy. {\bf 12}, 907 (2016).

\bibitem{Richerme2014}
P. Richerme, Z.-X. Gong, A. Lee, C. Senko, J. Smith, M. Foss-
Feig, S. Michalakis, A. V. Gorshkov, and C. Monroe, 
{\it Non-local propagation of correlations in quantum systems with long-range 
interactions},
Nature (London) {\bf 511}, 198 (2014).

\bibitem{Bernien2017}
H. Bernien, S. Schwartz, A. Keesling, H. Levine, A. Omran, 
H. Pichler, S. Choi, A. S. Zibrov, M. Endres, M. 
Greiner, V. Vuleti\'c, and M. D. Lukin, 
{\it Probing many-body dynamics on a 51-atom quantum simulator}, 
Nature {\bf 551}, 579 (2017). 

\bibitem{Periwal2021}
A. Periwal, E. S. Cooper, P. Kunkel, J. F. Wienand, E. J. Davis, and M. 
Schleier-Smith, 
{\it Programmable interactions and emergent geometry in an array of atom 
clouds}, 
Nature {\bf 600}, 630 (2021). 

\bibitem{Joshi2021}
M. K. Joshi, F. Kranzl, A. Schuckert, I. Lovas, C. Maier, R. Blatt, M. Knap, 
and 
C. F. Roos,
{\it Observing emergent hydrodynamics in a long-range quantum magnet}, 
Science {\bf 376}, 720 (2022).

\bibitem{Lieb1972}
E. H. Lieb and D. W. Robinson,
{\it The finite group velocity of quantum spin systems}, 
Commun. Math. Phys. {\bf 28}, 251 (1972).

\bibitem{Lashkari2013}
N. Lashkari, D. Stanford, M. Hastings, T. Osborne, and P. Hayden,
{\it Towards the fast scrambling conjecture}, 
J. High Energy Phys. {\bf 2013}, 22 (2013). 

\bibitem{Avellino2006}
M. Avellino, A. J. Fisher, and S. Bose, 
{\it Quantum communication in spin systems with long-range interactions}, 
Phys. Rev. A {\bf 74}, 012321 (2006). 

\bibitem{Hastings2006}
M. B. Hastings and T. Koma, 
{\it Spectral Gap and Exponential Decay of Correlations}, 
Commun. Math. Phys. {\bf 265}, 781 (2006).

\bibitem{Eisert2013}
J. Eisert, M. van den Worm, S. R. Manmana, and M. Kastner,
{\it Breakdown of Quasilocality in Long-Range Quantum Lattice Models}, 
Phys. Rev. Lett. {\bf 111}, 260401 (2013).

\bibitem{Hauke2013}
P. Hauke and L. Tagliacozzo, 
{\it Spread of Correlations in Long-Range Interacting Quantum Systems}, 
Phys. Rev. Lett. {\bf 111}, 207202 (2013).

\bibitem{Foss-Feig2015}
M. Foss-Feig, Z.-X. Gong, C. W. Clark, and A. V. Gorshkov,
{\it Nearly Linear Light Cones in Long-Range Interacting Quantum Systems}, 
Phys. Rev. Lett. {\bf 114}, 157201 (2015).

\bibitem{Tran2019}
M. C. Tran, A. Y. Guo, Y. Su, J. R. Garrison, Z. Eldredge, M.
Foss-Feig, A. M. Childs, and A. V. Gorshkov, 
{\it Locality and Digital Quantum Simulation of Power-Law Interactions}, 
Phys. Rev. X {\bf 9}, 031006 (2019).

\bibitem{Luitz2019}
D. J. Luitz and Y. Bar Lev, 
{\it Emergent locality in systems with power-law interactions}, 
Phys. Rev. A {\bf 99}, 010105(R) (2019). 

\bibitem{Colmenarez2020}
L. Colmenarez and D. J. Luitz, 
{\it Lieb-Robinson bounds and out-of-time order correlators in a long-range spin 
chain}, 
Phys. Rev. Research {\bf 2}, 043047 (2020). 

\bibitem{Zhou2019}
T. Zhou and X. Chen,
{\it Operator dynamics in a Brownian quantum circuit}, 
Phys. Rev. E {\bf 99}, 052212 (2019). 

\bibitem{Kuwahara2020}
T. Kuwahara and K. Saito, 
{\it Strictly Linear Light Cones in Long-Range Interacting Systems of Arbitrary 
Dimensions}, 
Phys. Rev. X {\bf 10}, 031010 (2020). 

\bibitem{Tran2021}
M. C. Tran, A. Y. Guo, C. L. Baldwin, A. Ehrenberg, 
A. V. Gorshkov, and A. Lucas, 
{\it Lieb-Robinson Light Cone for Power-Law Interactions}, 
Phys. Rev. Lett. {\bf 127}, 160401 (2021). 

\bibitem{Chen2021}
C.-F. Chen and A. Lucas, 
{\it Optimal Frobenius light cone in spin chains with power-law interactions}, 
Phys. Rev. A {\bf 104}, 062420 (2021). 

\bibitem{Kloss2019}
B. Kloss and Y. Bar Lev, 
{\it Spin transport in a long-range-interacting spin chain}, 
Phys. Rev. A {\bf 99}, 032114 (2019).

\bibitem{Schuckert2020}
A. Schuckert, I. Lovas, and M. Knap, 
{\it Nonlocal emergent hydrodynamics in a long-range quantum spin system}, 
Phys. Rev. B {\bf 101}, 020416 (2020). 

\bibitem{Cevolani2016}
L. Cevolani, G. Carleo, and L. Sanchez-Palencia, 
{\it Spreading of correlations in exactly solvable quantum models with 
long-range interactions in arbitrary dimensions}, 
New J. Phys. {\bf 18}, 093002 (2016).

\bibitem{Hashizume2022}
T. Hashizume, S. Kuriyattil, A. J. Daley, and G. Bentsen, 
{\it Tunable Geometries in Sparse Clifford Circuits}, 
arXiv:2202.11750. 

\bibitem{Schachenmayer2013}
J. Schachenmayer, B. P. Lanyon, C. F. Roos, and A. J. Daley,
{\it Entanglement Growth in Quench Dynamics with Variable Range Interactions}, 
Phys. Rev. X {\bf 3}, 031015 (2013). 

\bibitem{Pappalardi2018}
S. Pappalardi, A. Russomanno, B. \v{Z}unkovi\v{c}, F. Iemini, A. 
Silva, and R. Fazio, 
{\it Scrambling and entanglement spreading in long-range spin chains}, 
Phys. Rev. B {\bf 98}, 134303 (2018). 

\bibitem{Lerose2020}
A. Lerose and S. Pappalardi, 
{\it Origin of the slow growth of entanglement entropy in long-range interacting 
spin systems}, 
Phys. Rev. Research {\bf 2}, 012041(R) (2020). 

\bibitem{Kuwahara2021}
T. Kuwahara and K. Saito, 
{\it Absence of Fast Scrambling in Thermodynamically Stable Long-Range 
Interacting Systems}, 
Phys. Rev. Lett. {\bf 126}, 030604 (2021). 

\bibitem{Minato2022}
T. Minato, K. Sugimoto, T. Kuwahara, and K. Saito, 
{\it Fate of Measurement-Induced Phase Transition in Long-Range Interactions}, 
Phys. Rev. Lett. {\bf 128}, 010603 (2022). 

\bibitem{Mueller2022}
T. M\"uller, S. Diehl, and M. Buchhold, 
{\it Measurement-Induced Dark State Phase Transitions in Long-Ranged Fermion 
Systems}, 
Phys. Rev. Lett. {\bf 128}, 010605 (2022). 

\bibitem{Bachelard2013}
R. Bachelard and M. Kastner, 
{\it Universal Threshold for the Dynamical Behavior of Lattice Systems with 
Long-Range Interactions}, 
Phys. Rev. Lett. {\bf 110}, 170603 (2013).

\bibitem{Zaletel2015}
M. P. Zaletel, R. S. K. Mong, C. Karrasch, J. E. Moore, and F. Pollmann, 
{\it Time-evolving a matrix product state with long-ranged interactions}, 
Phys. Rev. B {\bf 91}, 165112 (2015).

\bibitem{Nielsen2000}
M. A. Nielsen and I. L. Chuang, {\it Quantum Computation and
Quantum Information} (Cambridge University Press, Cambridge, 2000).

\bibitem{Knill2008}
E. Knill, D. Leibfried, R. Reichle, J. Britton, R. B. Blakestad, J. D. Jost, C. 
Langer, R. Ozeri, S. Seidelin, and D. J. Wineland, 
{\it Randomized benchmarking of quantum gates}, 
Phys. Rev. A {\bf 77}, 012307 (2008). 

\bibitem{Magesan2011}
E. Magesan, J. M. Gambetta, and J. Emerson,
{\it Scalable and Robust Randomized Benchmarking of Quantum Processes}, 
Phys. Rev. Lett. {\bf 106}, 180504 (2011).

\bibitem{Li2018}
Y. Li, X. Chen, and M. P. A. Fisher, 
{\it Quantum Zeno effect and the many-body entanglement transition}, 
Phys. Rev. B {\bf 98}, 205136 (2018). 

\bibitem{Gullans2020}
M. J. Gullans and D. A. Huse,
{\it Dynamical Purification Phase Transition Induced by Quantum Measurements}, 
Phys. Rev. X {\bf 10}, 041020 (2020). 

\bibitem{Sharma2022}
S. Sharma, X. Turkeshi, R. Fazio, and M. Dalmonte, 
{\it Measurement-induced criticality in extended and long-range unitary 
circuits}, 
SciPost Phys. Core {\bf 5}, 023 (2022). 

\bibitem{Lavasani2021}
A. Lavasani, Y. Alavirad, and M. Barkeshli, 
{\it Measurement-induced topological entanglement transitions in symmetric 
random quantum circuits}, 
Nat. Phys. {\bf 17}, 342 (2021). 

\bibitem{Lunt2021_2}
O. Lunt, M. Szyniszewski, and A. Pal, 
{\it Measurement-induced criticality and entanglement clusters: A study of 
one-dimensional and two-dimensional Clifford circuits}, 
Phys. Rev. B {\bf 104}, 155111 (2021). 

\bibitem{Aaronson2004}
S. Aaronson and D. Gottesman,
{\it Improved simulation of stabilizer circuits}, 
Phys. Rev. A {\bf 70}, 052328 (2004).

\bibitem{Anders2006}
S. Anders and H. J. Briegel, 
{\it Fast simulation of stabilizer circuits using a graph-state 
representation}, 
Phys. Rev. A {\bf 73}, 022334 (2006). 

\bibitem{Gottesman1998}
D. Gottesman,
{\it The Heisenberg Representation of Quantum Computers}, 
arXiv:quant-ph/9807006

\bibitem{Note::Heisenberg}
Note that this expression differs from the usual Heisenberg picture, where the 
order of ${\cal U}$ and ${\cal U}^\dagger$ is reversed.

\bibitem{SuppMat}
See supplemental material for details on the structure of the Clifford group, 
additional numerical results on transport, operator spreading, and 
entanglement growth in circuits with and without U$(1)$ symmetry in one and 
two dimensions, and entanglement dynamics in a long-range 
Ising chain,   
including Refs.\ \cite{Corcoles2013, Koenig2014}. 


\bibitem{Corcoles2013}
A. D. C\'orcoles, J. M. Gambetta, J. M. Chow, J. A. Smolin, M. Ware, J. D.  
Strand, B. L. T. Plourde, and M. Steffen, 
{\it Process verification of two-qubit quantum gates by randomized benchmarking
}, 
Phys. Rev. A {\bf 87}, 030301(R) (2013). 

\bibitem{Koenig2014}
R. Koenig and J. A. Smolin, 
{\it How to efficiently select an arbitrary Clifford group element}, 
J. Math. Phys. {\bf 55}, 122202 (2014). 

\bibitem{Metzler2000}
R. Metzler and J. Klafter,
{\it The random walk's guide to anomalous diffusion: a fractional dynamics 
approach}, 
Phys. Rep. {\bf 339}, 1 (2000).

\bibitem{Zaburdaev2015}
V. Zaburdaev, S. Denisov, and J. Klafter, 
{\it L\'evy walks}, 
Rev. Mod. Phys. {\bf 87}, 483 (2015). 

\bibitem{Fattal2004}
D. Fattal, T. S. Cubitt, Y. Yamamoto, S. Bravyi, and I. L. Chuang, 
{\it Entanglement in the stabilizer formalism}, 
arXiv:quant-ph/0406168.

\bibitem{NoteZZ}
Note that if one were to simulate the dynamics of a state stabilized by $\pm 
Z_\ell$ on all lattice sites, i.e., a product state $\ket{\uparrow \downarrow 
\cdots }$ in the $Z$ basis, evolution with respect to $Z$-conserving Clifford 
gates would be entirely classical with $S(t) = 0$ for all $t$. This is in 
contrast to Haar-random gates with U$(1)$ symmetry.  

\bibitem{NoteAlpha}
We emphasize again that this corresponds to $1d$ Hamiltonian systems with 
couplings 
$J \propto r^{-\alpha'}$, with $\alpha'= \alpha/2$. 

\bibitem{Zanardi2001}
P. Zanardi, 
{\it Entanglement of quantum evolutions}, 
Phys. Rev. A {\bf 63}, 040304(R) (2001). 

\bibitem{Xu2022}
S. Xu and B. Swingle, 
{\it Scrambling Dynamics and Out-of-Time Ordered Correlators in Quantum 
Many-Body Systems: a Tutorial}, 
arXiv:2202.07060.

\bibitem{Hamma2005}
A. Hamma, R. Ionicioiu, and P. Zanardi, 
{\it Bipartite entanglement and entropic boundary law in lattice spin systems}, 
Phys. Rev. A {\bf 71}, 022315 (2005). 

\bibitem{Note_Mod2}
Note that the rank has to be calculated by treating the entries 
$\nu_\ell^i$,$\mu_\ell^i$ of ${\cal M}$ as elements of the field 
$\mathbb{F}_2$, 
i.e., addition and multiplication are performed modulo $2$. 

\bibitem{NoteLogCorrection}
In fact, we find the closest agreement with $z=2$ if we make the ansatz 
$S(t) \sim t^{1/z} \left(\log{t}\right)^{2/3}$, from which we obtain $z = 
1.98(1)$.

\bibitem{Leviatan2017}
E. Leviatan, F. Pollmann, J. H. Bardarson, D. A. Huse, and E. 
Altman, 
{\it Quantum thermalization dynamics with Matrix-Product States}, 
arXiv:1702.08894.

\bibitem{Ye2020}
B. Ye, F. Machado, C. D. White, R. S. K. Mong, and N. Y. Yao, 
{\it Emergent Hydrodynamics in Nonequilibrium Quantum Systems}, 
Phys. Rev. Lett. {\bf 125}, 030601 (2020). 

\bibitem{Kvorning2022}
T. K. Kvorning, L. Herviou, and J. H. Bardarson, 
{\it Time-evolution of local information: thermalization dynamics of local 
observables}, 
SciPost Phys. {\bf 13}, 080 (2022).

\bibitem{Keyserlingk2022}
C. W. von Keyserlingk, F. Pollmann, and T. Rakovszky,
{\it Operator backflow and the classical simulation of quantum transport}, 
Phys. Rev. B {\bf 105}, 245101 (2022).

\bibitem{Farshi2022}
T. Farshi, J. Richter, D. Toniolo, A. Pal, and L. Masanes,
{\it Absence of localization in two-dimensional Clifford circuits}, 
arXiv:2210.10129. 

\bibitem{Chandran2015}
A. Chandran and C. R. Laumann, 
{\it Semiclassical limit for the many-body localization transition},
Phys. Rev. B {\bf 92}, 024301 (2015). 

\bibitem{Agrawal2021}
U. Agrawal, A. Zabalo, K. Chen, J. H. Wilson, A. C. Potter, J. H. Pixley, S. 
Gopalakrishnan, and R. Vasseur, 
{\it Entanglement and charge-sharpening transitions in U(1) symmetric monitored 
quantum circuits}, 
arXiv:2107.10279. 

\bibitem{Bao2021}
Y. Bao, S. Choi, and E. Altman, 
{\it Symmetry enriched phases of quantum circuits}, 
Ann. Phys. {\bf 435}, 168618 (2021). 

\bibitem{Zhou2020_3}
S. Zhou, Z. Yang, A. Hamma, and C. Chamon,
{\it Single T gate in a Clifford circuit drives transition to universal 
entanglement spectrum statistics}, 
SciPost Phys. {\bf 9}, 087 (2020). 

\bibitem{Wei2021}
D. Wei, A. Rubio-Abadal, B. Ye, F. Machado, J. Kemp, K. Srakaew, 
S. Hollerith, J. Rui, S. Gopalakrishnan, N. Y. Yao, I. Bloch, and 
J. Zeiher, 
{\it Quantum gas microscopy of Kardar-Parisi-Zhang superdiffusion}, 
Science {\bf 376}, 716 (2022). 

\bibitem{Islam2015}
R. Islam, R. Ma, P. M. Preiss, M. Eric Tai, A. Lukin, M. Rispoli, and M. Greiner, 
{\it Measuring entanglement entropy in a quantum many-body system}, 
Nature {\bf 528}, 77 (2015).

\bibitem{Linke2018}
N. M. Linke, S. Johri, C. Figgatt, K. A. Landsman, A. Y. Matsuura, and C. 
Monroe, 
{\it Measuring the R\'enyi entropy of a two-site Fermi-Hubbard model on a 
trapped ion quantum computer}
Phys. Rev. A {\bf 98}, 052334 (2018). 


\end{thebibliography}

\begin{thebibliography}{99}

\bibitem{Corcoles2013S}
A. D. C\'orcoles, J. M. Gambetta, J. M. Chow, J. A. Smolin, M. Ware, J. D.  
Strand, B. L. T. Plourde, and M. Steffen, 
Phys. Rev. A {\bf 87}, 030301(R) (2013). 

\bibitem{Koenig2014S}
R. Koenig and J. A. Smolin, 
J. Math. Phys. {\bf 55}, 122202 (2014). 

\bibitem{Khemani2018S}
V. Khemani, A. Vishwanath, and D. A. Huse, 
Phys. Rev. X {\bf 8}, 031057 (2018). 

\bibitem{Zhou2020S}
T. Zhou, S. Xu, X. Chen, A. Guo, and B. Swingle,
Phys. Rev. Lett. 124, 180601 (2020). 

\bibitem{Schuckert2020S}
A. Schuckert, I. Lovas, and M. Knap, 
Phys. Rev. B {\bf 101}, 020416 (2020).

\bibitem{Rakovszky2019S}
T. Rakovszky, F. Pollmann, and C. W. von Keyserlingk,
Phys. Rev. Lett. {\bf 122}, 250602 (2019). 

\bibitem{Kim2013S}
H. Kim and D. A. Huse, 
Phys. Rev. Lett. {\bf 111}, 127205 (2013).

\bibitem{Block2022S}
M. Block, Y. Bao, S. Choi, E. Altman, and N. Y. Yao, 
Phys. Rev. Lett. {\bf 128}, 010604 (2022). 

 

\bibitem{Znidaric2020S}
M. \v{Z}nidari\v{c}, 
Commun. Phys. {\bf 3}, 100 (2020). 

\end{thebibliography}
 \end{document}